\documentclass{emulateapj}
\usepackage{apjfonts}

\slugcomment{To appear in the Astrophysical Journal, v659n1, April 2007}

\begin{document}

\title{A Multi-wavelength study of 1WGA~J1346.5$-$6255: a new $\gamma$~Cas 
analog unrelated to the background supernova remnant G309.2$-$00.6}

\author{Samar Safi-Harb\altaffilmark{1,2},
Marc Rib\'o\altaffilmark{3,4},
Yousaf Butt\altaffilmark{5},
Heather Matheson\altaffilmark{1},
Ignacio Negueruela\altaffilmark{6},
Fangjun Lu\altaffilmark{7},
Shumei Jia\altaffilmark{7},
Yong Chen\altaffilmark{7}}

\altaffiltext{1}{Department of Physics and Astronomy, University of Manitoba,
Winnipeg, MB, R3T 2N2, Canada; safiharb@cc.umanitoba.ca}
\altaffiltext{2}{Department of Physics, George Washington University, 725 21st Street NW, Washington, DC 20052}
\altaffiltext{3}{Departament d'Astronomia i Meteorologia, Universitat de 
Barcelona, Mart\'{\i} i Franqu\`es 1, 08028 Barcelona, Spain}
\altaffiltext{4}{AIM - Astrophysique Interactions Multi-\'echelles
(UMR 7158 CEA/CNRS/Universit\'e Paris 7 Denis Diderot),
CEA Saclay, DSM/DAPNIA/Service d'Astrophysique, B\^at. 709,
L'Orme des Merisiers, 91191 Gif-sur-Yvette Cedex, France}
\altaffiltext{5}{Harvard-Smithsonian Center for Astrophysics, 60 Garden St.,
Cambridge, MA 02138, USA}
\altaffiltext{6}{Departamento de F\'{\i}sica, Ingenier\'{\i}a de Sistemas y
Teor\'{\i}a de la Se\~nal, Escuela Polit\'ecnica Superior, Universitat
d'Alacant, Ap. 99, 03080 Alicante, Spain}
\altaffiltext{7}{Key Laboratory of Particle Astrophysics, Istitute of High 
Energy Physics, Chinese Academy of Sciences, Beijing 100049, P.R., China}

\shorttitle{1WGA~J1346.5$-$6255}
\shortauthors{Safi-Harb et al.}

\begin{abstract}

1WGA~J1346.5$-$6255 is a {\it ROSAT} X-ray source found within the radio 
lobes of the supernova remnant (SNR) G309.2$-$00.6. This source also 
appears to coincide with the bright and early-type star HD~119682, which 
is in the middle of the galactic open cluster NGC~5281. The radio 
morphology of the remnant, consisting of two brightened and distorted arcs 
of emission on opposite sides of the 1WGA~J1346.5$-$6255 source and of a 
jet-like feature and break in the shell, led to the suggestion that 
1WGA~J1346.5$-$6255/G309.2$-$00.6 is a young analog of the microquasar 
SS~433 powering the W50 nebula. This motivated us to study this source at 
X-ray and optical wavelengths. We here present new {\it Chandra} 
observations of 1WGA~J1346.5$-$6255, archival {\it XMM-Newton} 
observations of G309.2$-$00.6, and optical spectroscopic observations of 
HD~119682, in order to search for X-ray jets from 1WGA~J1346.5$-$6255, 
study its association with the SNR, and test for whether HD~119682 
represents its optical counterpart. We do not find evidence for jets from 
1WGA~J1346.5$-$6255 down to an unabsorbed flux of 2.6$\times$10$^{-13}$ 
ergs~cm$^{-2}$~s (0.5--7.5~keV), we rule out its association with 
G309.2$-$00.6, and we confirm that HD~119682 is its optical counterpart. 
We derive a distance of 1.2$\pm$0.3~kpc, which is consistent with the 
distance estimate to NGC~5281 (1.3$\pm$0.3~kpc), and much smaller than the 
distance derived to the SNR G309.2$-$00.6. We discuss the nature of the 
source, unveil that HD~119682 is a Be star and suggest it is a new member 
of the recently proposed group of $\gamma$-Cas analogs. The {\it Chandra} 
and {\it XMM-Newton} X-ray lightcurves show variability on timescales of 
hundreds of seconds, and the presence of a possible period of $\sim$1500~s 
that could be the rotational period of an accreting neutron star or white 
dwarf in this $\gamma$-Cas analog.

\end{abstract}


\section{Introduction}\label{introduction}

X-ray observations of Supernova Remnants (SNRs) have been continuously 
shaping our understanding of the diversity of compact objects associated 
with them. While the Crab used to be the prototype example for the 
aftermath of a supernova explosion, there is now growing evidence for a 
diversity of compact objects associated with SNRs. These include the 
magnetically powered neutron stars or magnetars, the radio-quiet neutron 
stars, the high-magnetic field radio pulsars, and accretion-powered 
candidates. Out of the 236 catalogued Galactic SNRs however, W50/SS~433 
remains the only system confirmed to harbor a microquasar (a neutron star 
or black hole in a binary system emitting a two-sided semi-relativistic 
jet). The search for other such systems is needed not only to address the 
uniqueness of W50/SS~433, but also to study the connection between 
supernovae and microquasars. To date, there are just six confirmed 
microquasars with massive optical companions in our Galaxy 
\citep{paredes05}. Since the physics of the accretion-ejection phenomenon 
in microquasars (and AGN) central engines is not well understood, each 
additional SNR-microquasar system helps greatly to constrain the 
theoretical models. Here we present a study of the 
G309.2$-$00.6/1WGA~J1346.5$-$6255 system which has been proposed to be a 
young analog of the W50/SS~433 system.

\subsection{G309.2$-$00.6 and 1WGA~J1346.5$-$6255} \label{gwga}

\begin{figure}
\begin{center}
\centerline {\plotone{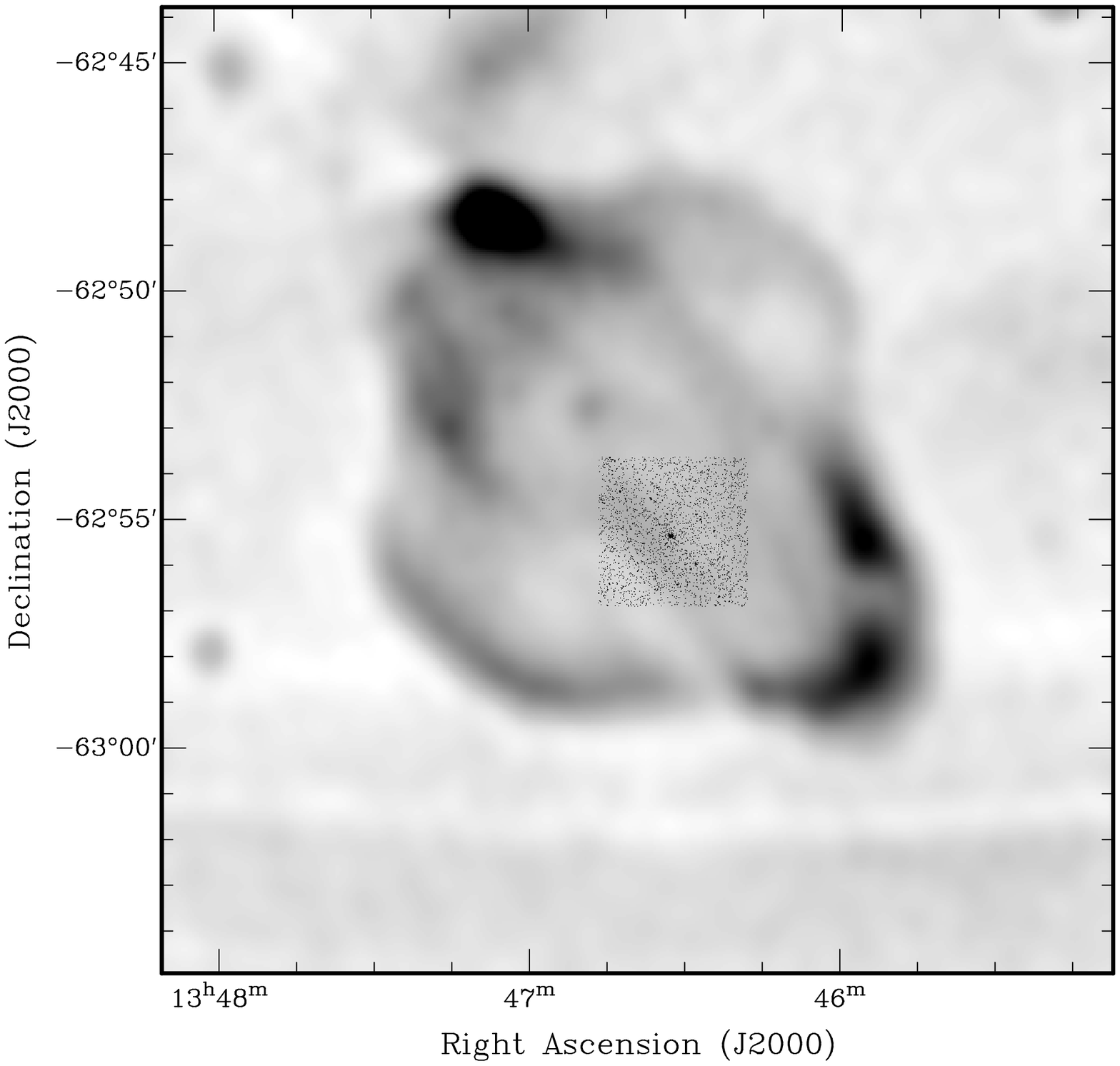}}
%\centerline {\plotone{f1-lowres.eps}}
%images/MOSTima-ChandraFOV-notitle.ps}}
\caption{{\it Chandra} image (inset) overlayed on to a radio image of the 
supernova remnant G309.2$-$00.6, spanning 21\arcmin$\times$21\arcmin, 
obtained with MOST at 0.843~GHz, with a resolution of 43\arcsec\ (Whiteoak 
\& Green 1996).}
\label{figure:G309_MOST_S3}
\end{center}
\end{figure}

\begin{figure}
\begin{center}
\centerline {\plotone{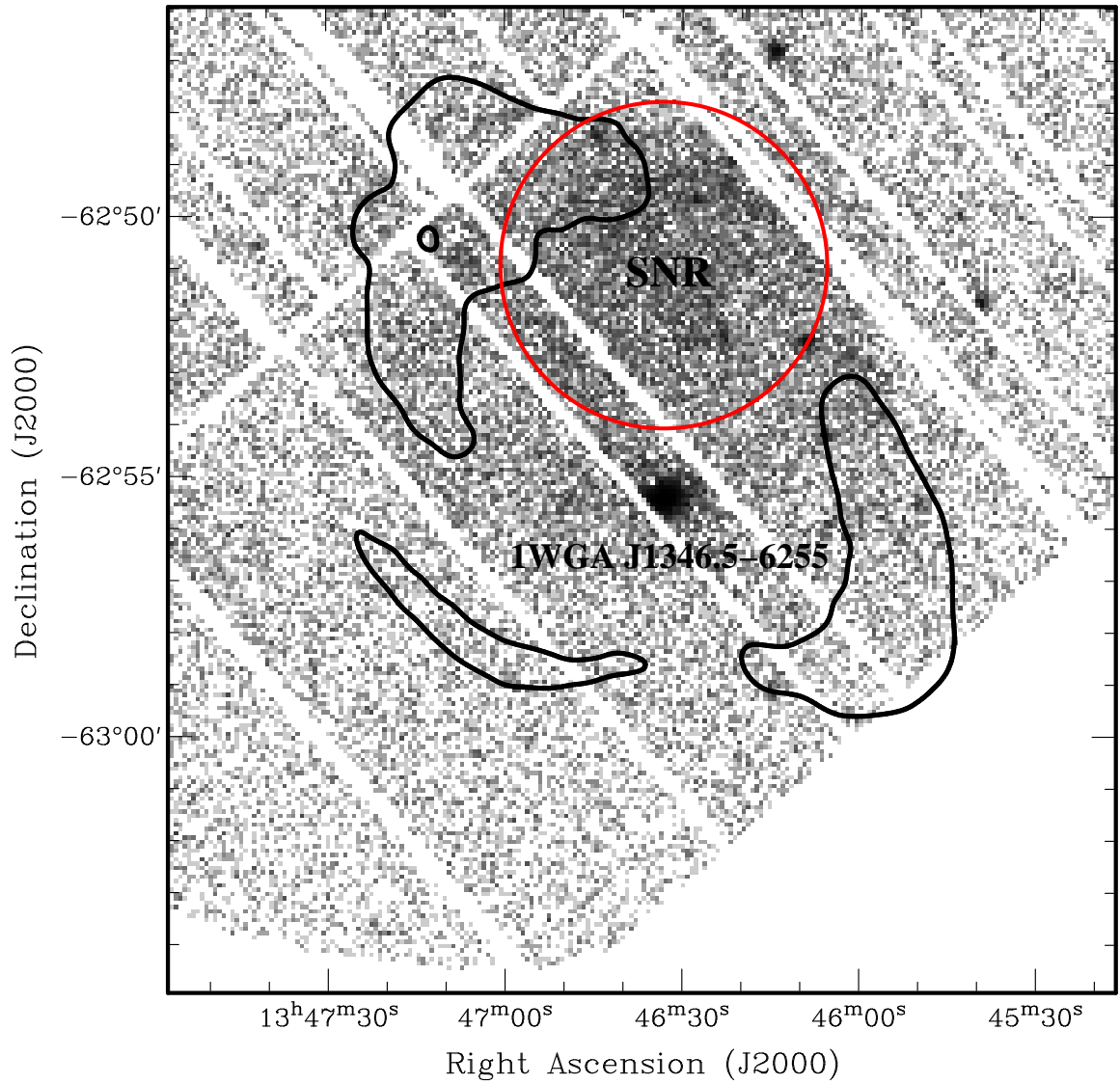}}
%\centerline {\plotone{f2-lowres.eps}}
%images/PNima-MOSTcont-SNRcirclecopy.xfig.eps}}
\caption{The {\it XMM-Newton}/PN image of 1WGA~J1346.5$-$6255 and 
G309.2$-$00.6, spanning 20\arcmin$\times$20\arcmin, shown in logarithmic 
scale. The overlaid black contour corresponds to the brightest emission 
from G309.2$-$00.6 detected at 0.843~GHz with MOST. The overlaid circle 
denotes the region extracted from the {\it XMM} data for the spectral 
analysis of the diffuse emission from the SNR (see \S4).}
\label{figure:G309_XMMima}
\end{center}
\end{figure}

G309.2$-$00.6 was first identified as an SNR based on its nonthermal radio 
emission \citep{green74}. A detailed study of the SNR was performed using 
the Australia Telescope Compact Array (ATCA, \citealt{gaensler98}). The 
remnant has a distorted shell-like morphology (see 
Fig.~\ref{figure:G309_MOST_S3} and Fig.~\ref{figure:G309_XMMima}) with two 
brightened and distorted arcs of emission on opposite sides with diameters 
of 12\arcmin\ and 15\arcmin, a jet-like feature and breaks in the shell. 
The SNR has a surface brightness of 5.4$\times$10$^{-21}$ 
W~m$^{-2}$~Hz$^{-1}$~Sr$^{-1}$ at 0.843~GHz \citep{whiteoak96}. 
Interestingly, this morphology resembles almost exactly a scaled down 
version of the SNR W50 (see, e.g., \citealt{dubner98}), well known to be 
powered by the microquasar SS~433. \ion{H}{1} absorption measurements 
against the SNR yield a distance in the range 5.4--14.1~kpc, corresponding 
to an age of (1--20)$\times$10$^3$~yr \citep{gaensler98}.

Infrared observations of the remnant show no obvious emission from 
G309.2$-$00.6 \citep{gaensler98}, but strong emission coincident with the 
nearby object RCW~80, identified as an \ion{H}{2} region based on its 
infrared and H$\alpha$ emission and flat radio spectrum. \ion{H}{1} 
absorption measurements indicate a distance of 5.4$\pm$1.6~kpc towards 
RCW~80. \cite{gaensler98} suggest that the SNR may be interacting with the 
adjacent \ion{H}{2} region, which would put the SNR at the lower range of 
the suggested distance and would imply an age $\leq$4,000~years. The 
striking resemblance of the radio morphology of G309.2$-$00.6 to W50 and 
the radio jet feature which aligns with the axis of symmetry defined by 
the two ears led \cite{gaensler98} to suggest that the SNR is powered by a 
central source producing collimated outflows, or jets, in two opposed 
directions, like the microquasar SS~433. A central source, however, has 
not been detected in the radio band, with a 5$\sigma$ upper limit of 
0.4~mJy at 1.3~GHz.

X-ray observations of G309.2$-$00.6 with the Advanced Satellite for 
Cosmology and Astrophysics ({\it ASCA}) have shown that the SNR is rich in 
thermal emission from a metal-rich ejecta-dominated plasma 
\citep{rakowski01}, indicating a young SNR age (700--4,000 years old). The 
X-ray spectrum of the SNR is best fitted with a non-equilibrium ionization 
model with a column density $N_{\rm 
H}\sim0.6\pm0.3\times10^{22}$~cm$^{-2}$, yielding an estimated distance of 
4$\pm$2~kpc to the remnant.

A bright {\it ROSAT } X-ray point-like source, 1RXS~J134633.6$-$625528 
also known as 1WGA~J1346.5$-$6255 \citep{white96}, with a count rate of 
0.06$\pm$0.02 counts~s$^{-1}$ is located near the center of the radio 
emission from the SNR. This source is coincident with the early-type star 
HD~119682, the third brightest star in the middle of the galactic open 
cluster NGC~5281 (see \S\ref{hd}). \cite{rakowski01} suggested the 
possibility that this source is related to the SNR based on its position 
and derived column density of (0.3--1.0)$\times$10$^{22}$~cm$^{-2}$.

\subsection{HD~119682} \label{hd}

The star HD~119682 has been usually classified as one of the members of 
NGC~5281. Although it has been observed several times (see available $V$ 
and $B-V$ photometry in Table~\ref{table:optical}) no precise spectral 
type and luminosity class have ever been reported, preventing an accurate 
estimation of the distance to this star. HD~119682 was initially 
classified as having a B spectral type by \cite{cannon18}, and later on 
\cite{stephenson71} gave an OB+ spectral type. \cite{moffat73} suggest a 
luminosity class of III? and quote a distance of 1.30~kpc to NGC~5281 
based on its unevolved stars. These authors also note that star \#3, 
HD~119682, has an unusual position in the color-color diagram (CCD) and 
color-magnitude diagram (CMD). \cite{humphreys75} reports a Be spectral 
type, while \cite{fitzgerald79} give a slightly different spectral type of 
O9e, although no luminosity class is specified. Based on the position of 
HD~119682 in the CCD and CMDs by \cite{moffat73}, \cite{mermilliod82} 
classified the object as a blue straggler.

A detailed study of NGC~5281 was carried out by \cite{sanner01}. They 
report further photometry for HD~119682 and explicitly state that they do 
not see evidence for a blue straggler nature in the CMDs when compared to 
other stars of NGC~5281. Based on isochrone fitting to the CMDs of the 
cluster members they provide the following parameters: 
$(m-M)_0=11.0\pm0.2$~mag, $d=1580\pm150$~pc, $E_{B-V}=0.20\pm0.02$~mag, an 
age of $t=45\pm10$~Myr (or $\log t ({\rm yr})=7.65\pm0.10$) and a solar 
metallicity of $Z=0.02$.

\cite{levenhagen04} report the following physical parameters by fitting 
the spectral profile of HD~119682 with a non-LTE model: $v \sin i = 
220\pm20$~km~s$^{-1}$, $T_{\rm eff}=31910\pm550$~K (or $\log T_{\rm 
eff}{\rm (K)}=4.504\pm0.008$), $\log g=4.00\pm0.10$~dex, $\log t ({\rm 
yr})=6.60\pm0.10$ (or $t=4\pm1$~Myr), $\log L/L_\odot=4.64\pm0.10$ and 
$M=18\pm1~M_\odot$. Note that the derived age is an order of magnitude 
smaller than the one of NGC~5281 reported by \cite{sanner01}, clearly 
reinforcing the blue straggler nature of the star.

Finally, we note that near infrared photometry of HD~119682 is present in 
the 2MASS catalog: $J=7.350\pm0.027$, $H=7.184\pm0.033$, and 
$K_s=6.950\pm0.023$~mag \citep{cutri03,skrutskie06}.

The goal of this paper is to study the nature of the 1WGA~J1346.5$-$6255 
source, its association with the optical star HD~119682, and the 
association of either source with the SNR G309.2$-$00.6. The previous 
proposal that G309.2$-$00.6 represents a young analog of W50 motivated us 
to study the {\it ROSAT} X-ray source and any associated jets with the 
{\it Chandra} X-ray Observatory. We also complement our X-ray study with 
archival observations obtained with {\it XMM-Newton}. While a detailed 
study of the diffuse emission from the SNR is beyond the scope of the 
paper, we here focus on the possible association between the X-ray source 
and the SNR. We also present new optical observations of HD~11968 acquired 
with ESO's New Technology Telescope. These observations are targeted to 
pin down the spectral type and luminosity class of the star from which to 
obtain a good estimate of its distance.

The paper is organized as follows. In \S\ref{observations}, we summarize 
the observations undertaken at X-ray and optical wavelengths. In 
\S\ref{chandra}, \ref{xmm}, and \ref{optical}, we present our {\it 
Chandra}, {\it XMM-Newton}, and the optical results, respectively. We 
finally discuss our results in \S\ref{discussion} and summarize our 
conclusions in \S\ref{summary}.

\section{Observations and data reduction} \label{observations}

The {\it ROSAT} source 1WGA~J1346.5$-$6255 was observed with the {\it 
Chandra} X-Ray Observatory for 15.12~ks on 2004 December 26, with the 
back-illuminated chip S3 of the Advanced CCD Imaging Spectrometer (ACIS-S; 
G. Garmire\footnote{See http://cxc.harvard.edu/proposer/POG}), at a focal 
plane temperature of $-$120~$^\circ$C~\footnote{For a review of {\it 
Chandra}, see 
http://cxc.harvard.edu/cdo/about$_{-}$chandra/overview$_{-}$cxo.html}. The 
data was processed using the standard CIAO~3.2.1 
tools\footnote{http://cxc.harvard.edu/ciao}. Events with ASCA grades (0, 
2, 3, 4, 6) were retained, and periods of high-background rates were 
removed. The net effective exposure time was 14.9~ks. The data was 
corrected for Charge Transfer Inefficiency (CTI) using the Penn State 
corrector \citep{townsley02} and filtered using the filtering available in 
acis\_process\_events (CIAO 3.2.1 with CALDB version 3.0.1).

The 1WGA~J1346.5$-$6255 source spectrum was created by selecting photons 
within a circle of radius 3\farcs5 and centered at the peak of the X-ray 
emission (see \S3.1). The background was selected from an annulus centered 
on the X-ray source, with inner radius of 10\arcsec\ and outer radius of 
15\arcsec. The count rate for the source was $0.104\pm0.005$ 
counts~sec$^{-1}$ and the count rate for the background was 
$0.004\pm0.002$ counts~sec$^{-1}$. The spectrum was grouped to a minimum 
of 20 counts per bin and the energy range selected for spectral fitting 
was 0.5--7.5~keV (the spectrum is background dominated above 7.5~keV). The 
spectral analysis was performed using XSPEC version 11.2~\footnote 
{http://xspec.gsfc.nasa.gov}.

The SNR G309.2$-$00.6 was observed by the European Photon Imaging Camera 
(EPIC) aboard {\it XMM-Newton} \citep{jansen01} for 40.6~ks on 2001 August 
28 (obs ID 0087940201). The EPIC-PN camera was operating in Extended Full 
Frame mode, and the EPIC-MOS cameras in Full Frame mode. A thick filter 
was used to avoid problems with bright stars in the open cluster NGC~5281. 
The data were processed using version 6.0 of the {\it XMM-Newton} Science 
Analysis Software (SAS). We created filtered event files using the EPIC-PN 
and MOS cameras in the 0.3--10~keV energy range. We then selected events 
for which the pattern was less than or equal to 4 (12) for PN (MOS), we 
removed the bad pixels, and cleaned the data by removing the times with 
high background proton flares. The resulting effective exposure time is 
23.9~ks for PN, 27.8~ks for MOS1, and 27~ks for MOS2. The spectrum for 
1WGA~J1346.5$-$6255 was extracted using a circle centered at the peak of 
X-ray emission: $\alpha=13^{\rm h}~46^{\rm m}~32\fs6$, 
$\delta=-62\degr~55\arcmin~27\arcsec$ (J2000) and with a 0\farcm5 radius. 
The spectrum of the diffuse emission from G309.2$-$00.6 was extracted 
using a circle centered at the peak of the diffuse X-ray emission: 
$\alpha=13^{\rm h}~46^{\rm m}~33^{\rm s}$, 
$\delta=-62\degr~50\arcmin~57\arcsec$ (J2000) and a radius of 3\farcm125. 
This region is shown in Fig.~\ref{figure:G309_XMMima}. The background 
spectrum for 1WGA~J1346.5$-$6255 was extracted from a ring centered at the 
source and extending from 0\farcm5 to 1\farcm0; while the background 
spectrum for the diffuse emission was extracted from an annulus centered 
at the diffuse emission and of 5\farcm83 and 9\farcm17 for the inner and 
outer radii, respectively. The ancillary and response matrix files were 
produced using the SAS commands {\it arfgen} and {\it rmfgen}, 
respectively. The spectra were subsequently grouped into a minimum of 50 
counts per bin for the point source and 200 counts per bin for the diffuse 
emission. The spectral analysis was subsequently performed using XSPEC 
v11.2.

We observed HD~119682 using the ESO Multi-Mode Instrument (EMMI) on the 
3.5-m New Technology Telescope (NTT) at La Silla, Chile, on 2003 June 5th. 
An H$\alpha$ spectrum was taken with the red arm in 
intermediate-resolution mode (REMD) and grating \#6. The red arm is 
equipped with a mosaic of two thin, back-illuminated $2048\times4096$ 
MIT/LL CCDs and this configuration results in a nominal dispersion of 
0.4~\AA~pixel$^{-1}$ over the range $\lambda\lambda$~6440--7150~\AA. The 
resolution, measured on arc lines, is $\approx 1.2$~\AA. The blue spectrum 
was taken with the blue arm in intermediate-resolution mode (BLMD) and 
grating \#12. The blue arm is equipped with a Textronik TK1034 thinned, 
back-illuminated $1024\times1024$ CCD and this configuration results in a 
nominal dispersion of 0.9~\AA~pixel$^{-1}$ over the range 
$\lambda\lambda$~3820--4750~\AA. The resolution, measured on arc lines, is 
$\approx 2.6$~\AA. Image pre-processing was carried out with {\em MIDAS} 
software, while data reduction was achieved with the {\em Starlink} 
packages {\sc ccdpack} \citep{draper00} and {\sc figaro} 
\citep{shortridge97}.

\section{{\it Chandra} results} \label{chandra}

\subsection{Imaging} \label{imaging}

In Fig.~\ref{figure:G309_MOST_S3}, we show the radio image of the SNR 
G309.2$-$00.6 obtained with the Molonglo Observatory Synthesis Telescope 
(MOST) at a frequency of 0.843~GHz overlayed with the X-ray emission 
detected in the S3 chip of {\it Chandra}. The peak of the X-ray emission 
is located at $\alpha=13^{\rm h}~46^{\rm m}~32\fs6$, 
$\delta=-62\degr~55\arcmin~24\arcsec$ (J2000) with an error radius of 
1\farcs3 (90\% confidence level). This source corresponds to the 
previously known {\it ROSAT} source 1WGA~J1346.5$-$6255.

\begin{figure}
\begin{center}
\centerline {\plotone{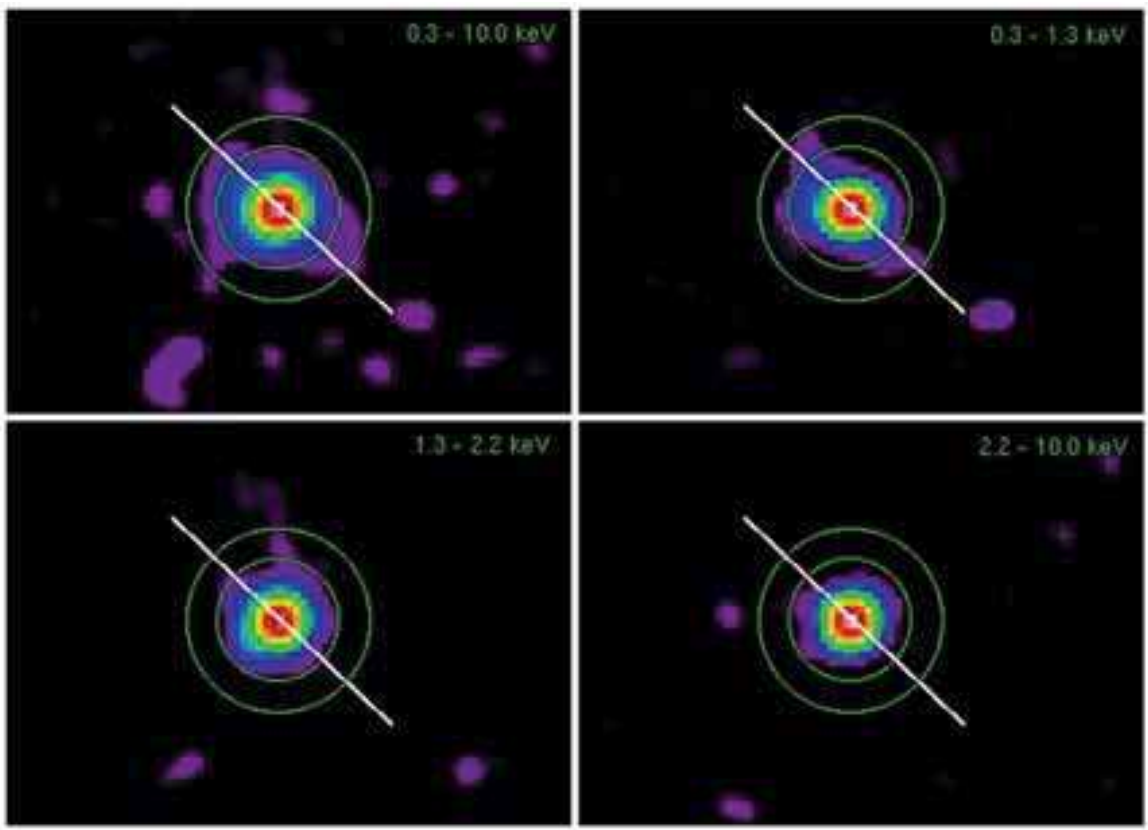}}
%\centerline {\plotone{f3-lowres.eps}}
%images/penn_smooth_4plots_wlines.eps}}
\caption{Top left: {\it Chandra} image of 1WGA~J1346.5$-$6255 in the 
0.3--10~keV energy range. The three other images correspond to the soft 
(top right), medium (bottom left) and hard (bottom right) energy ranges. 
The concentric circles have radii of 2 (barely visible), 4, and 6 
arcseconds. The white line indicates the axis of symmetry of the 
G309.2$-$00.6 radio lobes. the images have been smoothed (minimum 
smoothing scale = 1 pixel, minimum signal-to-noise ratio = 2). There is a 
hint of extension in the soft energy band along the symmetry axis of the 
radio lobes, although the low count rate indicates that it is most likely 
background noise.}
\label{figure:smoothed}
\end{center}
\end{figure}

Our primary goal is to search for any jets associated with this X-ray 
source. Since jets from compact objects generally peak at different X-ray 
energies than their powering engines, we created energy images for 
1WGA~J1346.5$-$6255 in the soft (0.3--1.3~keV), medium (1.3--2.2~keV), and 
hard (2.2--10~keV) band. The energy boundaries were chosen as to obtain a 
similar number of counts ($\sim$500) in each image. The images were then 
smoothed using the CIAO tool {\it csmooth} with a circular Gaussian kernel 
and are shown in Fig.~\ref{figure:smoothed}. The initial smoothing scale 
was 1 pixel. The smoothing scale was increased (to a maximum 2 pixels) 
until the number of counts under the kernel exceeded a signal-to-noise 
ratio of 2. Where the signal-to-noise ratio was less than 2, the image was 
smoothed on a scale of 2 pixels. The axis overlayed on top of the images, 
and running from northeast to southwest, delineates the axis of symmetry 
of the radio lobes of G309.2$-$00.6, $\simeq45\degr$ counter-clockwise 
from north \citep{gaensler98}. Interestingly, there is a hint of extension 
in the soft energy band along this axis. However, any extension observed 
in Fig.~\ref{figure:smoothed} (purple, dark blue) has a count rate of less 
than 0.7 counts~pixel$^{-1}$, indicating that the apparent extension is 
most likely background noise.

\begin{figure}
\begin{center}
\centerline {\plotone{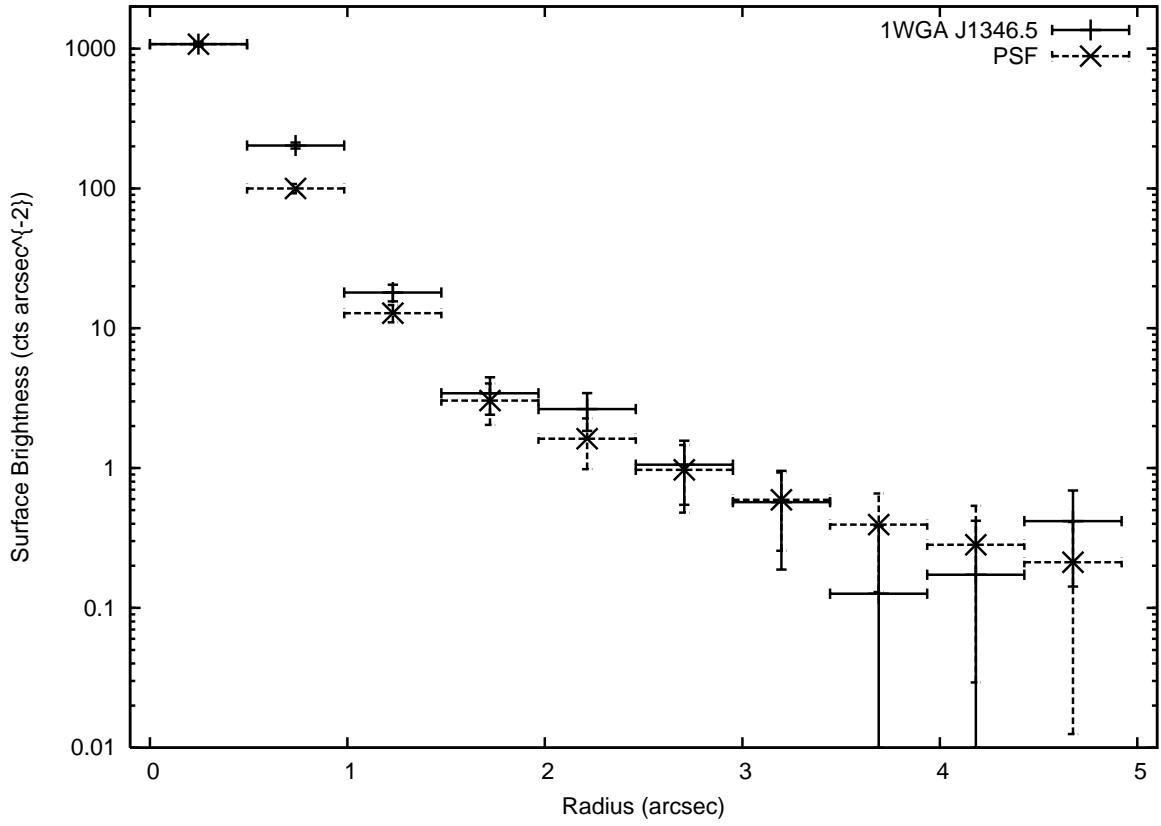}}
%images/radial_profile.ps}}
\caption{Radial profile of 1WGA~J1346.5$-$6255 (solid bars with 90\% 
confidence errors) and of {\it Chandra}'s PSF (dashed bars) in the 
0.3--10.0~keV range (see \S\ref{imaging} for details). The data are
consistent with a point-like source.}
\label{figure:all_data_psf}
\end{center}
\end{figure}

In order to further determine whether 1WGA~J1346.5$-$6255 is point-like or 
extended, we subsequently compared its spatial characteristics with {\it 
Chandra}'s point spread function (PSF). We generated a PSF image at an 
offset axis of 0\farcm48 (the location of the source on the S3 chip) and 
at an energy of 1.5~keV (characteristic of the source's energy histogram). 
We subsequently normalized the PSF image to the source counts ($\sim$1450 
counts in the 0.3--10~keV range) and used the PSF as a convolution kernel 
when fitting the source. Fitting a two-dimensional Gaussian distribution 
to the data, we found that the full width half maximum (FWHM) was only 
slightly larger than the FWHM of the PSF which is assumed to be circular 
(1\farcs02 and 0\farcs73, respectively). The data and PSF are shown in 
Fig.~\ref{figure:all_data_psf}. Since the difference is less than one 
pixel ($0\farcs4920\pm0\farcs0001$), the data is consistent with a point 
source. A similar profile was created for the soft (0.3--2.4~keV) and hard 
energy band (2.4--10~keV), by evaluating the PSF at 1.1~keV and 2.5~keV, 
respectively. The FWHM of the data and PSF in the soft band was found to 
be 0\farcs97 and 0\farcs73, respectively. For the hard energy band, the 
FWHM of the data and PSF was 1\farcs09 and 0\farcs77, respectively. We 
conclude that the data in either band are consistent with a point source, 
and therefore rule out the presence of jets associated with 
1WGA~J1346.5$-$6255, down to an unabsorbed flux of 
2.6$\times$10$^{-13}$~ergs~cm$^{-2}$~s$^{-1}$ in the 0.5--7.5~keV range 
(see next section).

\subsection{Spectroscopy} \label{spectroscopy}

\begin{figure}
\begin{center}
{\includegraphics[width=12cm]{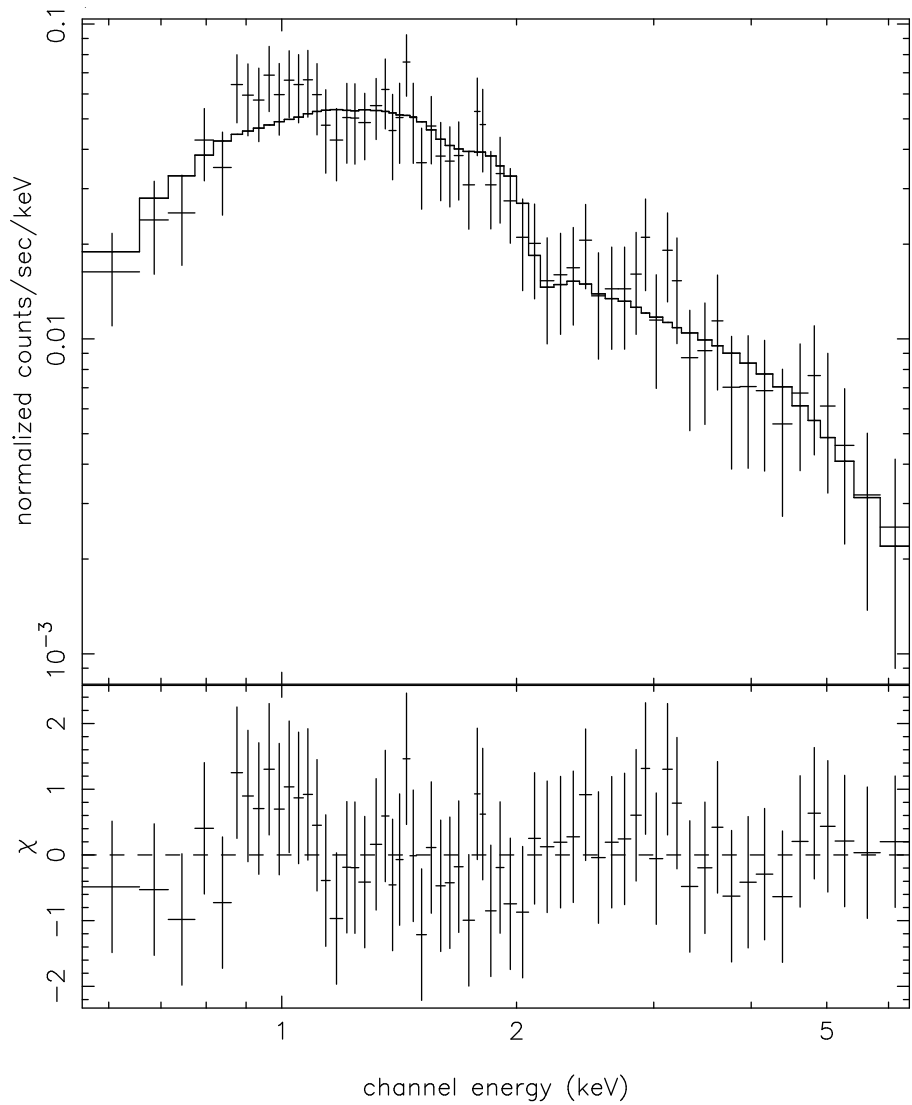}}
%images/point_source_CHANDRA_pl_ld.xfig.eps}
\caption{The {\it Chandra} spectrum of the point source 
1WGA~J1346.5$-$6255 fitted with an absorbed power-law model. The ratio of 
data to fitted model is shown in the bottom panel. An excess is seen 
around 1~keV.}
\label{figure:Chandra_powerlaw_fit}
\end{center}
\end{figure}

In order to address the nature of 1WGA~J1346.5$-$6255, we fitted its 
spectrum with thermal and non-thermal models, modified by interstellar 
absorption. In Table~\ref{table:chandra}, we summarize the parameters of 
the best fit models. Errors are at the 90\% confidence level throughout 
the paper, unless otherwise mentioned. We find that blackbody and thermal 
bremsstrahlung models do not provide acceptable fits. The absorbed 
power-law model provides an adequate fit with the following parameters: a 
column density $N_{\rm H}=0.18^{+0.08}_{-0.07}\times10^{22}$~cm$^{-2}$, a 
photon index $\Gamma=1.4\pm0.2$ and an unabsorbed flux of 
$1.1^{+0.3}_{-0.2}\times$10$^{-12}$~ergs~cm$^{-2}$~s$^{-1}$ in the 
0.5--7.5~keV range (reduced $\chi^2_{\nu}$=0.46, $\nu$=61 degrees of 
freedom). The spectrum and fitted model are shown in 
Fig.~\ref{figure:Chandra_powerlaw_fit}, while in 
Fig.~\ref{figure:Chandra_powerlaw_cont} we show the 68, 95 and 99.7\% 
confidence contours for $N_{\rm H}$ and $\Gamma$. By examining the 
residuals in Fig.~\ref{figure:Chandra_powerlaw_fit}, some line emission 
appears to exist around 1~keV. For this purpose, we added a MEKAL 
component (which would, e.g., account for any emission associated with 
coronal emission from a normal star, see \S\ref{nature} for more details). 
We find that the fit improves significantly ($\chi^2_{\nu}$ = 0.341, $\nu$ 
= 59), with an F-test probability of 4.5$\times 10^{-5}$ that this 
improvement occurs by chance. This best fit model is shown in 
Fig.~\ref{figure:powmek} and the parameters are: $N_{\rm H}$ = 
0.19$^{+0.13}_{-0.08}$ $\times$ 10$^{22}$ cm$^{-2}$, $\Gamma$ = 
1.24$^{+0.28}_{-0.26}$, $kT$ = 0.97$^{+0.37}_{-0.28}$~keV, with an 
unabsorbed flux in the 0.5--7.5~keV energy range of 
$1.18^{+0.35}_{-0.29}\times$10$^{-12}$~ergs~cm$^{-2}$~s$^{-1}$, very 
similar to the one obtained with a pure absorbed power law.

In order to test whether the X-ray emission is coronal emission from a 
normal star (see \S\ref{nature}), pure MEKAL models were attempted. A 
single temperature MEKAL model yields a statistically acceptable fit, 
however the temperature is high and unconstrained (see 
Table~\ref{table:chandra}). A two-component MEKAL model yields as good a 
fit as the power-law+MEKAL model but with the temperature of the hard 
component poorly constrained: $N_{\rm H}$ = 0.19$^{+0.09}_{-0.07}$ 
$\times$ 10$^{22}$ cm$^{-2}$, $kT_1$ = 0.95$^{+0.34}_{-0.24}$~keV; $kT_2$ 
= 80 ($\geq$14)~keV, and a reduced $\chi^2_{\nu}$ = 0.342 for $\nu$ = 59 
degrees of freedom. The temperature of the hard component, of 
$\sim9\times10^8$~K, is unrealistically high and poorly constrained--we 
show in the next section that the {\it XMM-Newton} data helps better 
constrain this component. We conclude that while the two-component MEKAL
model and the power-law+MEKAL model both yield statistically acceptable fits,
the {\it Chandra} spectrum is best fitted and its 
parameters better constrained using the power-law+MEKAL 
model.

In order to quantify the upper limit on the flux from any unseen jets 
associated with 1WGA~J1346.5$-$6255, we added a power-law component to the 
best fit power-law+MEKAL model above. We then froze the parameters of the
model for 1WGA~J1346.5$-$6255 and allow the photon index for any `unseen'
jets to vary between 1.6--2.0,
which is reasonable based on observations of other
jet sources. We find that the corresponding upper limit
on the unabsorbed flux is 2.6$\times$10$^{-13}$~ergs~cm$^{-2}$~s$^{-1}$
in the 0.5--7.5 keV range.
At a distance of 1.3~kpc (see \S6), this   
translates to a luminosity of $\sim$5.3$\times$10$^{31}$~ergs~s$^{-1}$.
% We froze the parameters of the model the 
%for 1WGA~J1346.5$-$6255 and allowed the photon index parameter for any the 
%`unseen' jets to vary between 1.0 and 2.0 (which is reasonable for jets the 
%associated with compact objects). We find that this three-component the 
%power-law+power-law+MEKAL model fit the data as long as the flux from any the 
%underlying jets is $\leq$2.6$\times$10$^{-13}$~ergs~cm$^{-2}$~s$^{-1}$ the 
%(unabsorbed, 0.5--7.5~keV). At a distance of 1.3~kpc (see \S6), this the 
%translates to a luminosity of $\sim$5.3$\times$10$^{31}$~ergs~s$^{-1}$.

\subsection{Timing} \label{timing}

To check for variability of 1WGA~J1346.5$-$6255, we created lightcurves 
with {\it Chandra} data binned at 100, 120 and 150~s time intervals. The 
obtained lightcurves display variability with an amplitude at the 
$\pm$30\% level (1$\sigma$ value) on a timescale of a few 100 seconds.

We have searched for periodic signals using standard techniques like the Phase
Dispersion Minimization (PDM, \citealt{stellingwerf78}) and the CLEAN algorithm
\citep{roberts87}. The PDM periodograms obtained for the {\it Chandra} data,
binned at either 100, 120 or 150~s, reveal a periodicity of
1500$^{+100}_{-50}$~s. The obtained PDM statistic $\Theta$ is in the range
0.89--0.92. An $F$-test reveals that this periodicity could be the result of a
random fluctuation with a probability of $\sim$40\% (see details in
\citealt{stellingwerf78}). This is not strange given the fact that the noise
witihin individual PDM phase bins is two times higher than the overall detected
variability. We note that simulations with even lower signal-to-noise ratios
have revealed the good performance of PDM (as can be seen for the primary
periodicity in \citealt{otazu02,otazu04}). On the other hand, two additional
possible periodicities of $\sim$3050 and $\sim$4600~s are detected with lower
significance. Given their spectral profiles, it is clear that they are
subharmonics of the $\sim$1500~s periodicity (see \citealt{stellingwerf78} for
details). The CLEAN algorithm displays less significant maxima at
$\sim$1550~s, possibly because the signal is not perfectly sinusoidal but is
better accounted for by a saw-tooth pattern with a slow increase and a
relatively fast decay. However, sinusoidal fits to the binned data provide a
possible period in the range 1545--1555~s, with a variability amplitude of
16\%. We show in Fig.~\ref{figure:lc_chandra} the background subtracted {\it
Chandra} lightcurve binned at 150~s intervals and normalized to its average
value. We also present the same data smoothed with a running window of 450~s
to enhance the 1500~s modulation, and a sinusoidal fit to the binned data, 
which approximately follows the relative maxima and minima of the smoothed 
data. In \S6.5, we discuss these results, together with the light curves 
obtained using the {\it XMM-Newton} data (next section).

\section{{\it XMM-Newton} results} \label{xmm}

\subsection{1WGA~J1346.5$-$6255}

In order to fit the {\it XMM-Newton} spectrum of 1WGA~J1346.5$-$6255, we 
attempted single component (thermal and non-thermal) and two-component 
models, as we did above for the {\it Chandra} data (\S\ref{spectroscopy}). 
The {\it XMM} data allowed us however to extend the fits up to 8.5~keV. As 
with {\it Chandra}, we found that the power-law model provides an adequate 
fit with $N_{\rm H}=0.22\pm0.02 \times 10^{22}$~cm$^{-2}$, a photon index 
$\Gamma$=1.69$^{+0.04}_{-0.05}$, and an unabsorbed X-ray flux of 
$2.22\pm0.02 \times 10^{-12}$~ergs~cm$^{-2}$~s$^{-1}$ in the 0.5--8.5~keV 
energy range, with a reduced $\chi^2_{\nu}$=1.04 (for $\nu$=418 degrees of 
freedom). As with the {\it Chandra} section, errors are at the 90\% 
confidence level, unless otherwise mentioned. The spectrum is shown in 
Fig.~\ref{figure:XMM_powerlaw_fit}, and the confidence contours for 
$N_{\rm H}$ and $\Gamma$ are shown in Fig.~\ref{figure:XMM_powerlaw_cont}.

\begin{figure}
\begin{center}
\centerline {\plotone{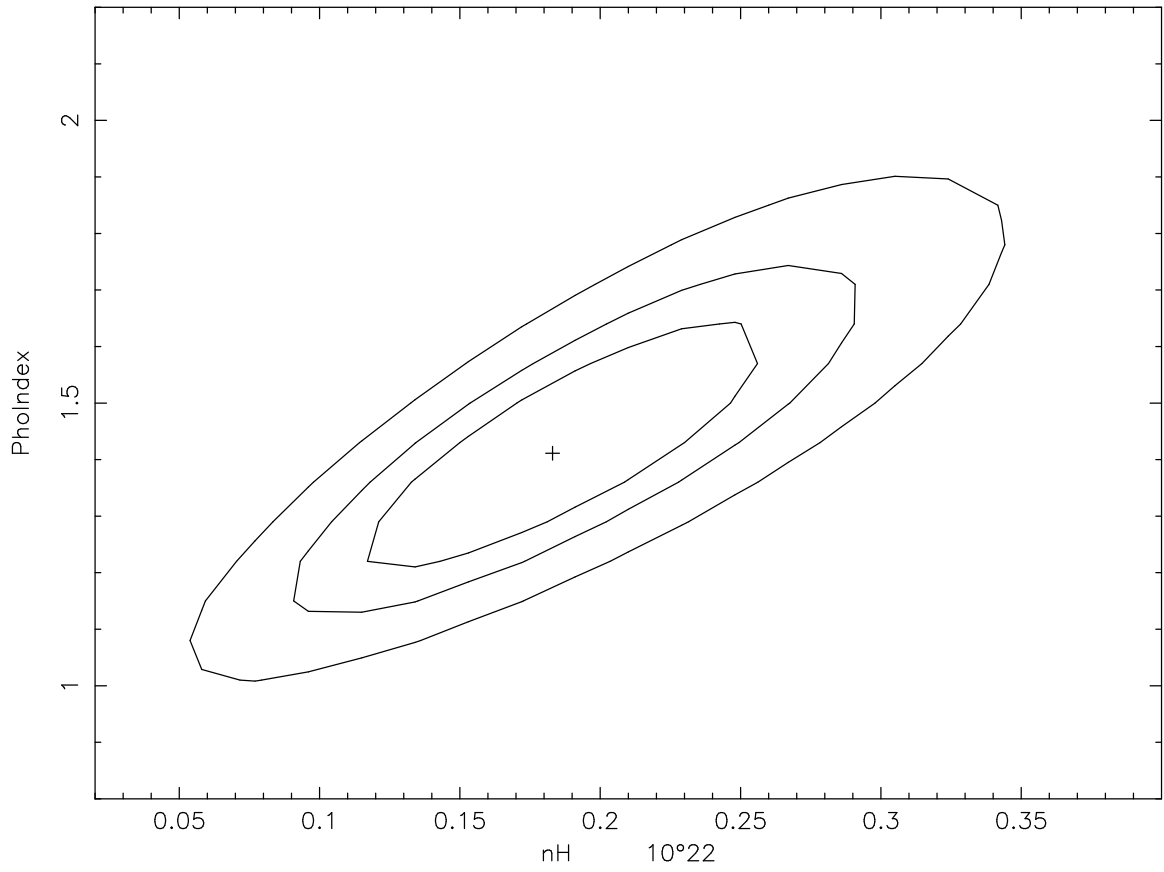}}
%images/power_cont_CHANDRA.xfig.eps}}
\caption{The 68, 95 and 99.7\% confidence contours for the absorbed
power-law model fit to the {\it Chandra} spectrum of 1WGA~J1346.5$-$6255.}
\label{figure:Chandra_powerlaw_cont}
\end{center}
\end{figure}

\begin{figure}
\begin{center}
{\includegraphics[width=12cm]{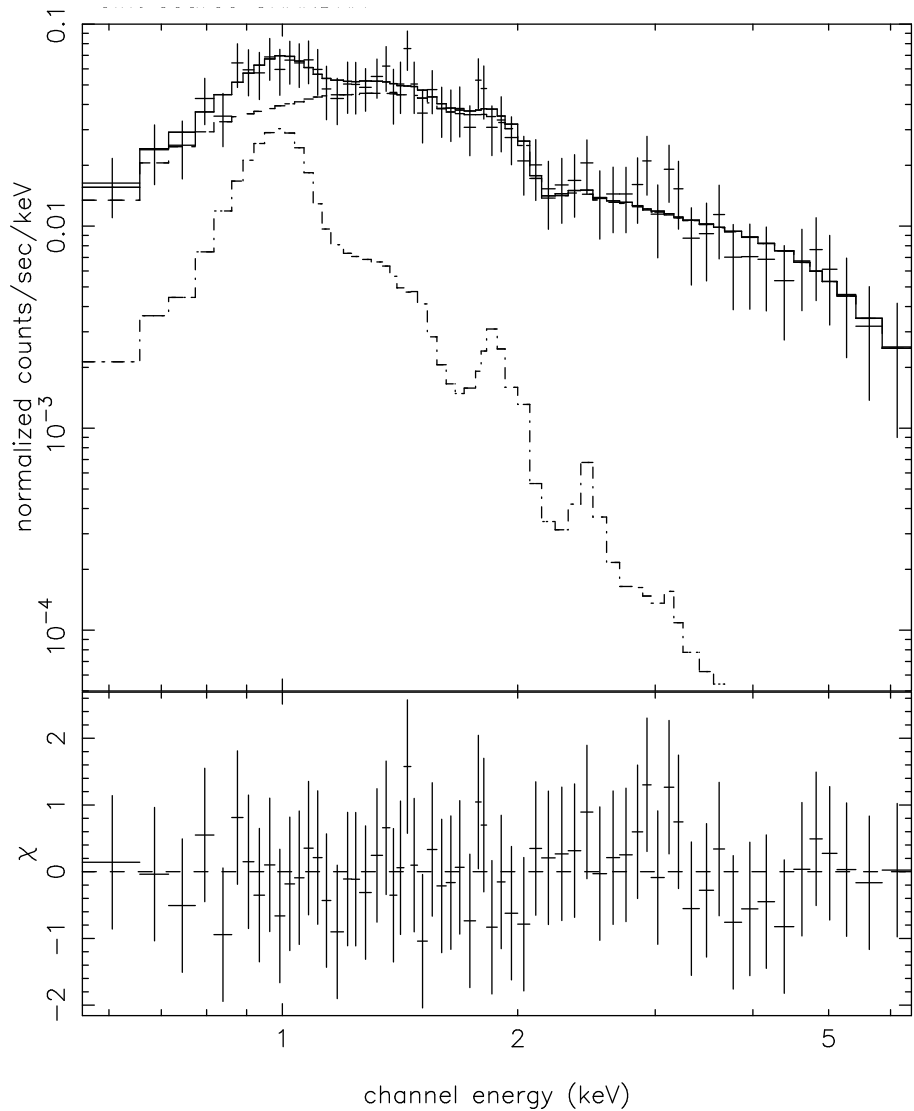}}
%images/mekal+pl.xfig.eps}}
\caption{The {\it Chandra} spectrum of the point source 
1WGA~J1346.5$-$6255 fitted with an absorbed power-law+MEKAL model. The 
dot-dashed line represents the contribution from the MEKAL soft component, 
while the dashed line is for the harder power-law component. The ratio of 
data to fitted model is shown in the bottom panel.}
\label{figure:powmek}
\end{center}
\end{figure}

\begin{figure}
\begin{center}
\centerline {\plotone{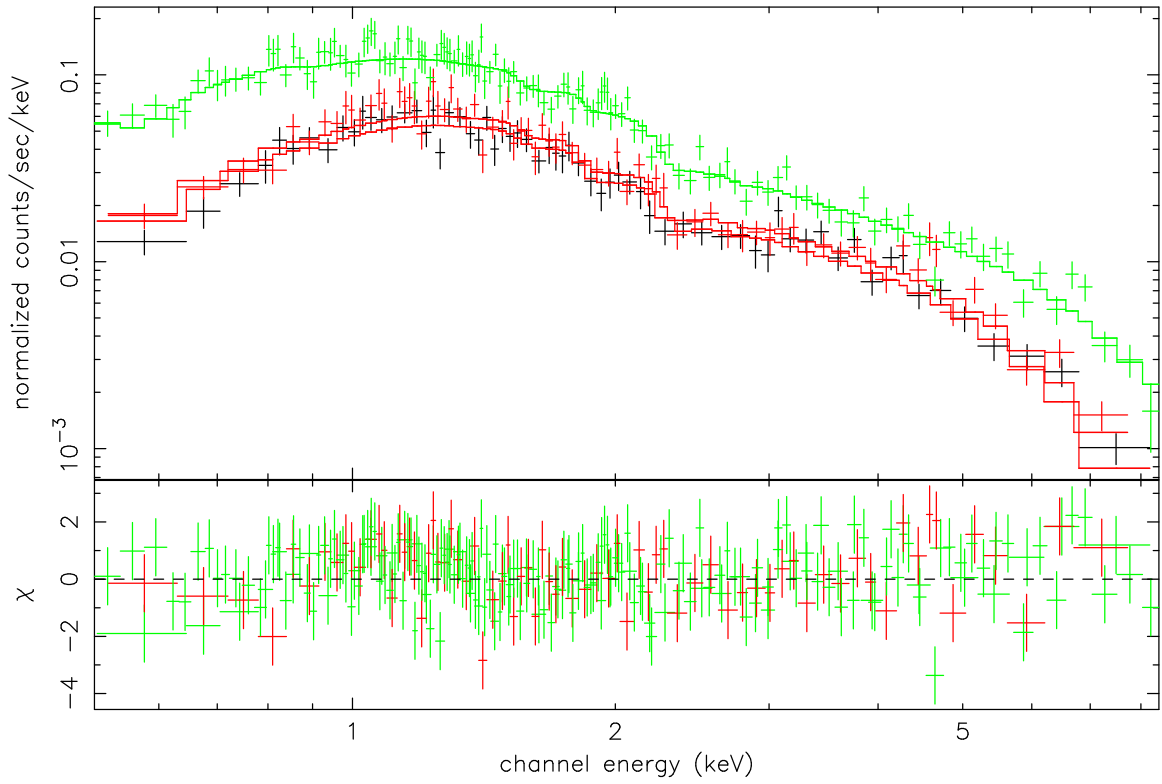}}
%images/ps_m1m2pn_pl.xfig.eps}}
\caption{The {\it XMM-Newton} PN (green), MOS1 (black) and MOS2 (red) 
spectra of of the point source 1WGA~J1346.5$-$6255 fitted with an absorbed 
power-law model. The bottom panel displays the ratio of data to fitted 
model.}
\label{figure:XMM_powerlaw_fit}
\end{center}
\end{figure}

\begin{figure}
\begin{center}
\centerline {\plotone{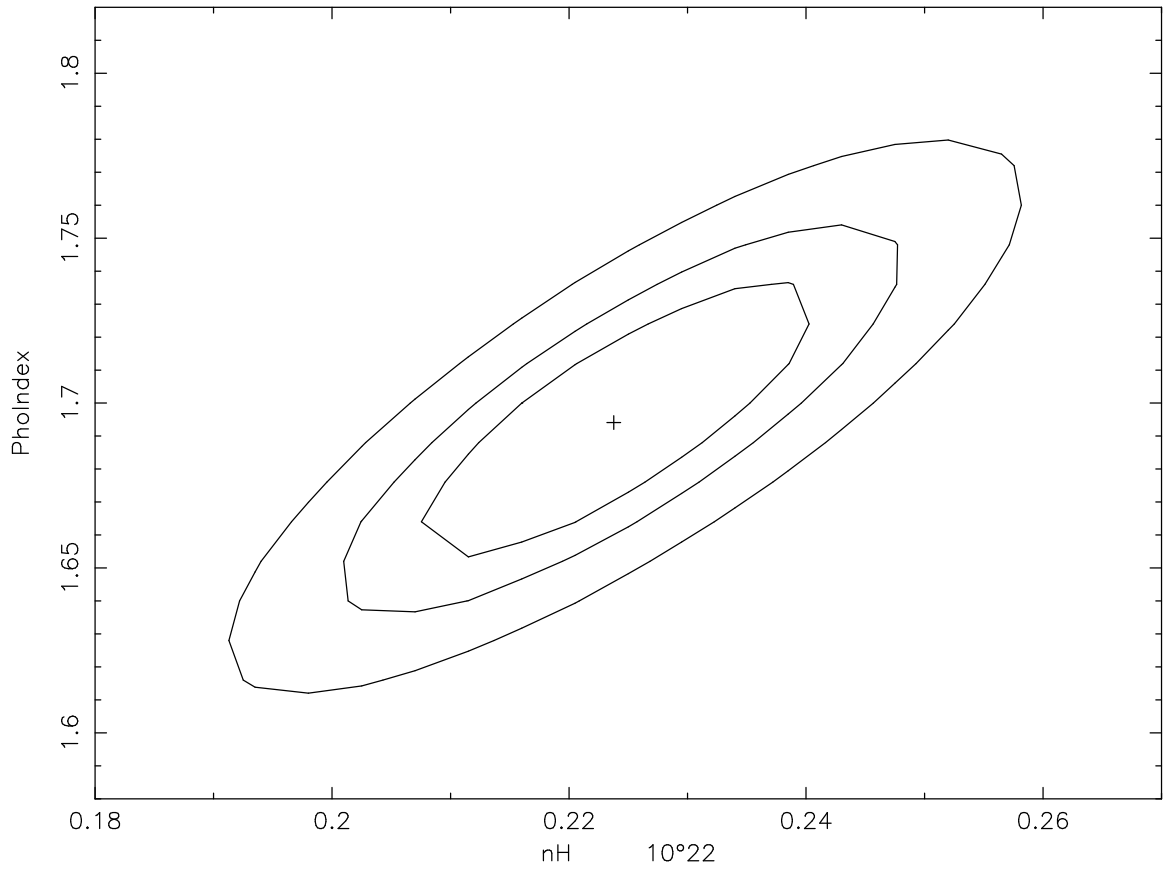}}
%images/mos1mos2pn-power_cont.xfig.eps}}
\caption{The 68, 95 and 99.7\% confidence contours for the absorbed
power-law model fit to the {\it XMM-Newton} PN and MOS spectra of
1WGA~J1346.5$-$6255.}
\label{figure:XMM_powerlaw_cont}
\end{center}
\end{figure}

By comparing the {\it XMM-Newton} to the {\it Chandra} parameters, we note 
that the {\it XMM}-derived $N_{\rm H}$ and $\Gamma$ values are slightly 
higher than those derived from the {\it Chandra} analysis 
(\S\ref{spectroscopy}). The contour plots however 
(Figs.~\ref{figure:Chandra_powerlaw_cont} and 
\ref{figure:XMM_powerlaw_cont}) show that within errors, the values are 
consistent with each other, with the {\it XMM-Newton} values being much 
better constrained.

We subsequently tested for whether pile-up affected the {\it Chandra} 
spectral analysis (the {\it XMM-Newton} spectrum is not affected by 
pile-up). Using Webpimms and a count rate of $\sim$0.10 counts~s$^{-1}$ in 
the 0.5--7.5~keV, we estimate a pile-up fraction of $\leq$12\% in ACIS-S3. 
We then corrected for pile-up using the {\it pileup} model in XSPEC. We 
find that the $N_{\rm H}$ and $\Gamma$ values increase slightly and are 
closer to the best values obtained with {\it XMM-Newton}, but the range is 
still consistent with our previous {\it Chandra} estimates and the contour 
levels are very similar to those shown in 
Fig.~\ref{figure:Chandra_powerlaw_cont}. After correction for pile-up, we 
find that $N_{\rm H}$ = 0.20$^{+0.07}_{-0.09}$ $\times10^{22}$~cm$^{-2}$, 
and $\Gamma$ = 1.53$^{+0.25}_{-0.33}$. The corresponding unabsorbed flux 
is $1.1^{+0.4}_{-0.2}\times10^{-12}$~ergs~cm$^{-2}$~s$^{-1}$, consistent 
with the value before pile-up correction, but only accounting for 
$\sim$50--55\% of the flux detected by {\it XMM-Newton} in the 
0.5--7.5~keV range. We conclude that while the spectral parameters are not 
significantly affected by pile-up, the difference in the flux estimate 
between the {\it Chandra} and {\it XMM} observations may be partly 
attributed to some contamination by the background and likely evidence for 
variability in the source (see \S\ref{discussion} for a discussion on the 
variability of the source).

\begin{figure}
\begin{center}
\centerline {\plotone{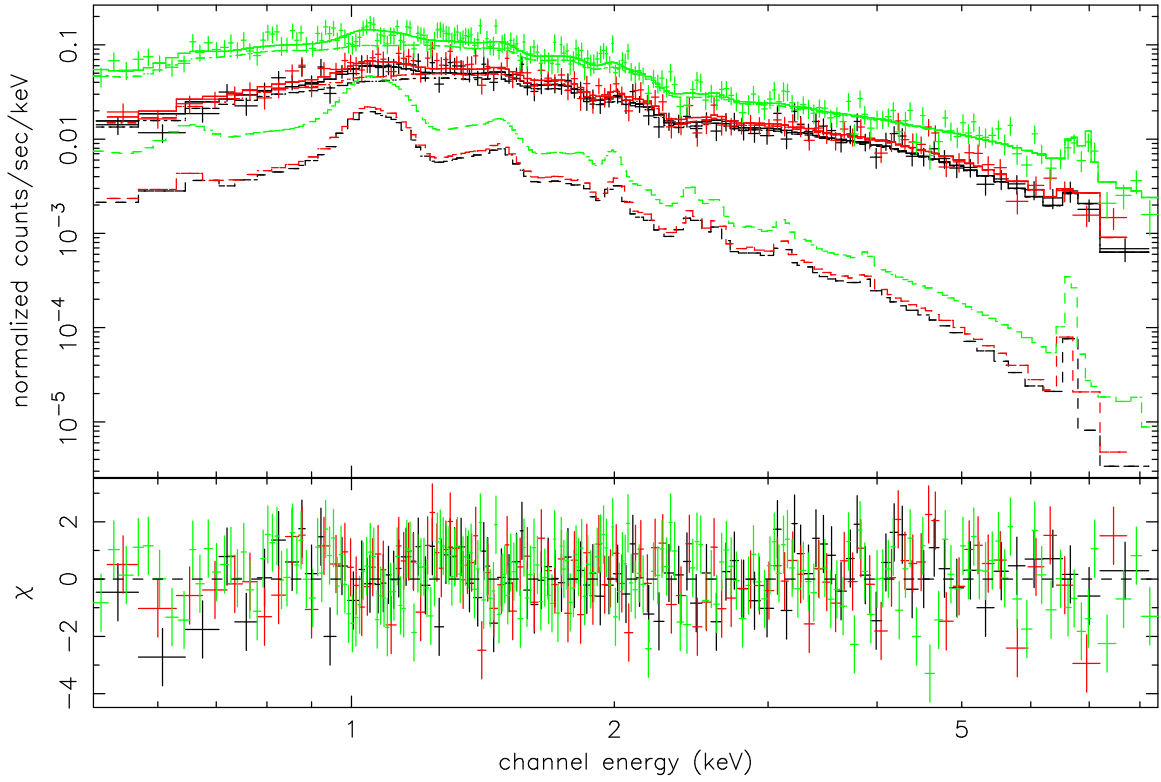}}
%images/ps_m1m2pn_mekal+mekal.xfig.eps}}
\caption{The {\it XMM-Newton} PN (green), MOS1 (black) and MOS2 (red) 
spectra of 1WGA~J1346.5$-$6255 fitted with a two-component absorbed MEKAL
model. The lower dashed lines indicate the contribution from the softer 
component (see text for details). The bottom panel displays the ratio of
data to fitted model.}
\label{figure:XMM_2mekal}
\end{center}
\end{figure}

\begin{figure}
\begin{center}
\centerline {\plotone{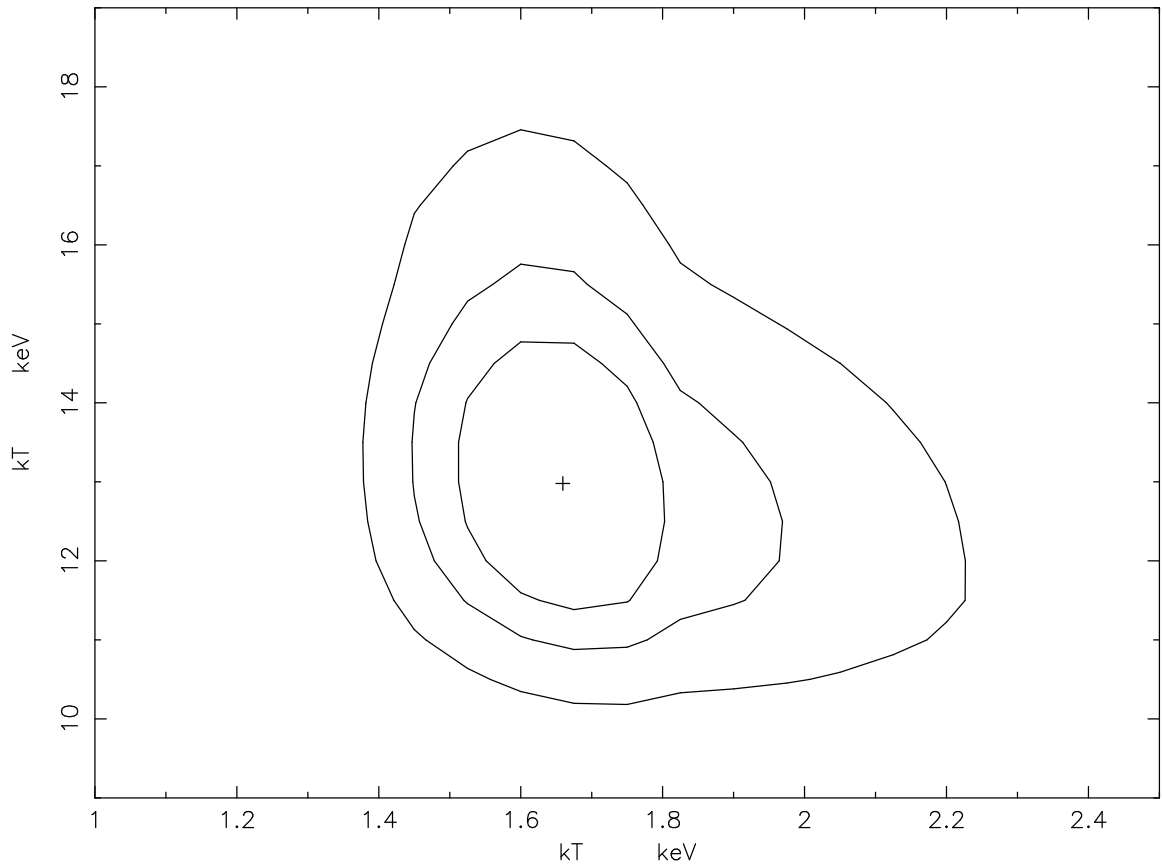}}
%images/ps_m1m2pn_mekal+mekal_kTscont.xfig.eps}} 
\caption{The 68, 95 and 99.7\% confidence contours for the absorbed 
two-component MEKAL model fit to the {\it XMM-Newton} PN and MOS spectra 
of 1WGA~J1346.5$-$6255.}
\label{figure:XMM_2mekal_cont}
\end{center}
\end{figure}

We then attempted pure MEKAL models. A one-temperature component MEKAL 
model provides a statistically acceptable fit, however the model 
overestimates the X-ray emission between 6--7~keV and is therefore 
discarded. Just like with the {\it Chandra} data, a two-temperature MEKAL 
model provides a better fit with the following parameters: $N_{\rm H}$= 
0.16$\pm$0.01 $\times$10$^{22}$~cm$^{-2}$, $kT_1$= 1.7$\pm$0.3~keV, 
$kT_2$= 13.0$^{+2.6}_{-2.4}$~keV, $\chi^2_{\nu}$ = 1.025 ($\nu$=416 
degrees of freedom). In Fig.~\ref{figure:XMM_2mekal}, we show the EPIC 
spectra fitted with this two-component model and in Table~\ref{table:xmm}, 
we summarize the results indicating that the harder component dominates 
the X-ray emission. In Fig.~\ref{figure:XMM_2mekal_cont}, we show the 68, 
95 and 99.7\% confidence contours for the fitted temperatures of the 
two-component MEKAL model. Note that the harder component's temperature is 
much better constrained with {\it XMM-Newton} than with {\it Chandra}. 
This temperature is still high for coronal emission but not inconsistent 
with the ones measured in the new $\gamma$~Cas-like objects (see the 
discussion section).

\begin{figure}
\begin{center}
\centerline {\plotone{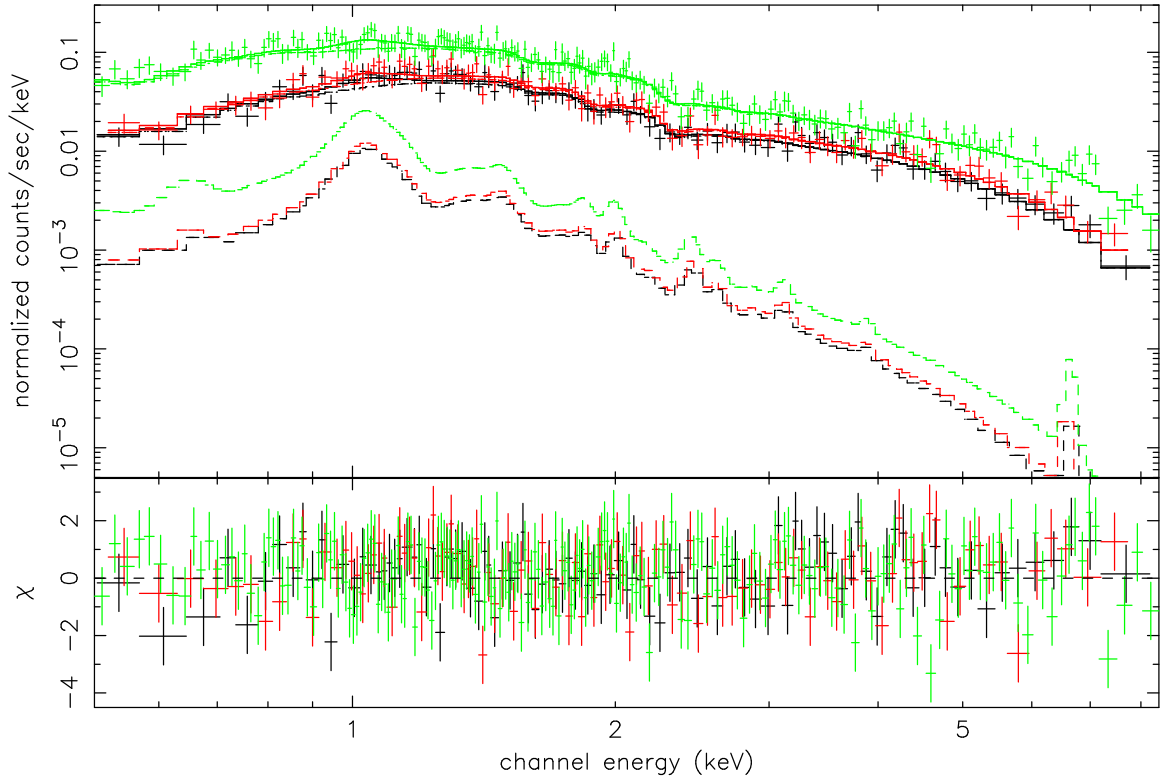}}
%images/ps_m1m2pn_pl+mekal.xfig.eps}} 
\caption{The {\it XMM-Newton} PN (green), MOS1 (black) and MOS2 (red) 
spectra of 1WGA~J1346.5$-$6255 fitted with a two-component absorbed 
power-law+MEKAL model. The lower dashed lines indicate the contribution 
from the softer (MEKAL) component. The bottom panel displays the ratio of 
data to fitted model.}
\label{figure:XMM_pl+mekal}
\end{center}
\end{figure}

Lastly, a two-component power-law+MEKAL model
provides the best fit with the
lowest $\chi^2_{\nu}$, with the 
following parameters: $N_{\rm H}$ = 
0.21$^{+0.03}_{-0.02}$$\times$10$^{22}$ cm$^{-2}$, $kT$ = 
1.4$^{+0.8}_{-0.3}$~keV, $\Gamma$ = 1.60$\pm$0.07, reduced 
$\chi^2_{\nu}$=1.005 ($\nu$=416 degrees of freedom). The corresponding fit 
is shown in Fig.~\ref{figure:XMM_pl+mekal} and the parameters summarized 
in Table~\ref{table:xmm}. As in the MEKAL+MEKAL fit, the harder component 
(now the power-law) dominates the X-ray emission.

Finally, it is worth noting that the power-law fit (see 
Fig.~\ref{figure:XMM_powerlaw_fit}) indicates that a bump near 6--7~keV is 
unaccounted for. Adding a Gaussian line to the power-law component 
improves the fit ($\chi^2_{415}$=1.012), with an F-test probability of 
0.117 that this improvement occurs by chance. The line energy parameters 
are: $E_{\rm line}$ = 6.77$^{+0.14}_{-0.12}$~keV, $\sigma_{\rm 
line}$=0.19$^{+0.13}_{-0.184}$~keV, and an unabsorbed line flux of 4.8 
$\times$ 10$^{-14}$ ergs~cm$^{-2}$~s$^{-1}$. Adding a Gaussian line near 
6.7~keV to the two-component (MEKAL+MEKAL or MEKAL+power-law) models does 
not improve the fits significantly.

We conclude that the two-component MEKAL+MEKAL, power-law+MEKAL, and 
power-law+Gaussian line, all provide good fits to the {\it XMM-Newton} 
data, although the power-law+MEKAL provides the best fit, a result that is 
consistent with the {\it Chandra} study.

As for the {\it Chandra} data, we also obtained background subtracted 
lightcurves binned at 150~s and normalized to their average values. As can 
be seen in Fig.~\ref{figure:lc_xmm}, 1WGA~J1346.5$-$6255 was significantly 
variable, with an amplitude at the $\pm$45--55\% level (1$\sigma$ values) 
on timescales of a few 100 seconds. A timing analysis reveals different 
possible signals at $\sim$4570, $\sim$8300 and $\sim$12000~s in the 
different lightcurves and using PDM or CLEAN. However, sinusoidal fits to 
the data with such periods do not converge. Since a marginal signal is 
nearly always detected at $\sim$1500~s with PDM and/or CLEAN, we have 
tried a sinusoidal fit around this periodicity, and found that convergence 
is achieved for a period of 1485$\pm$10~s and amplitudes of 10, 20 and 
16\% in the PN, MOS1 and MOS2 lightcurves (to be compared to the 16\% 
amplitude in the {\it Chandra} data, \S3.3). We have also analysed a 
combined data set of all normalized XMM data, and found similar results: 
the three possible periods reported above and a more clear signal than in 
individual data sets at 1486$\pm$2~s (and 2 subharmonics for PDM), with an amplitude of variability at 
the 13\% level, hence closely matching the {\it Chandra} results. 
An $F$-test to the PDM results reveals a probability of up to 60\% of 
this signal as being due to a random fluctuation. 
Although from the statistical point of view we can not argue that the periodicity exists, 
the detection of 2 subharmonics and the fact that we detect a very
similar periodicity with {\it Chandra} and {\it XMM} data gives us confidence that the period, 
or quasi-period, is real. Future observations can probe this issue.

\subsection{G309.2$-$00.6}

\begin{figure}
\begin{center}
\centerline {\plotone{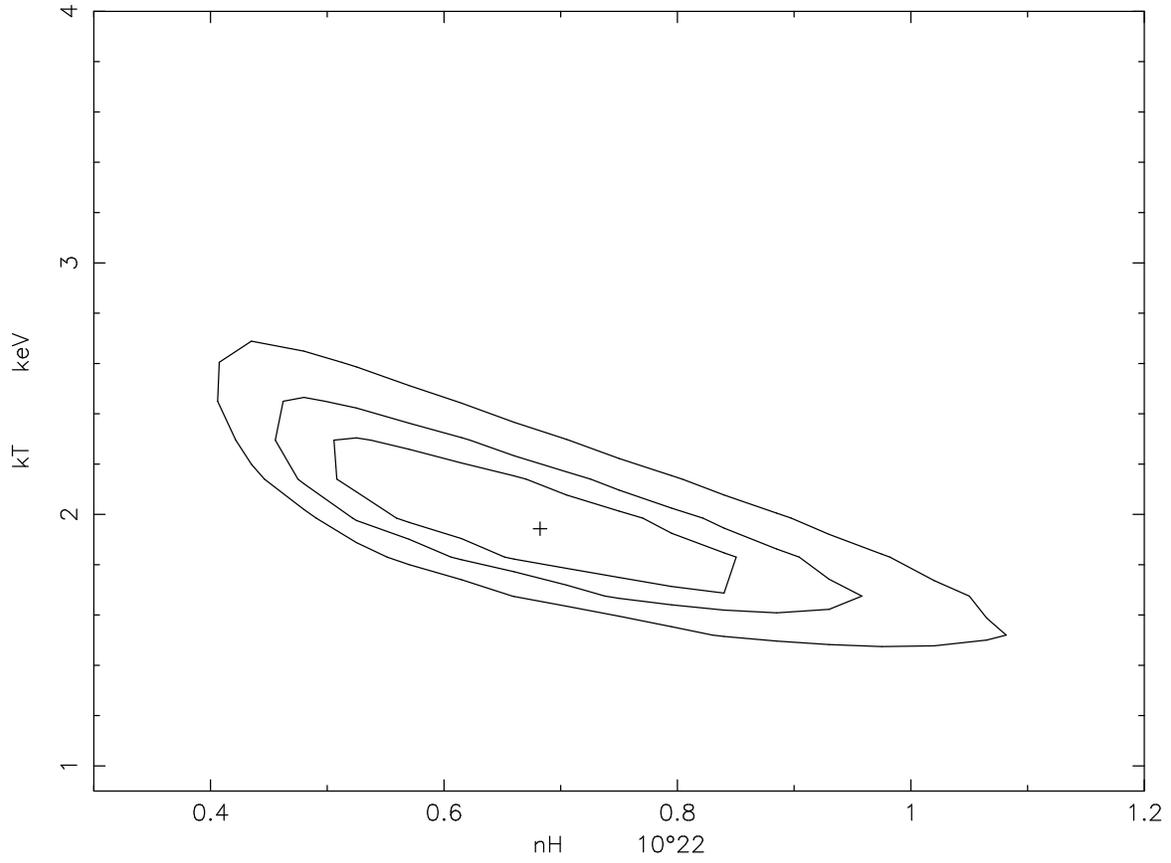}}
%images/snr_m1m2pn_vnei_vonemgsisarcafeni_NHkTcont_portrait.xfig.eps}}
\caption{The 68, 95 and 99.7\% confidence contours for the absorbed {\it 
vnei} model fit to the {\it XMM-Newton} PN spectrum of the diffuse 
emission from G309.2$-$00.6.}
\label{figure:snr_m1m2pn_vnei_NHkTcont}
\end{center}
\end{figure}

\begin{figure}
\begin{center}
\centerline {\plotone{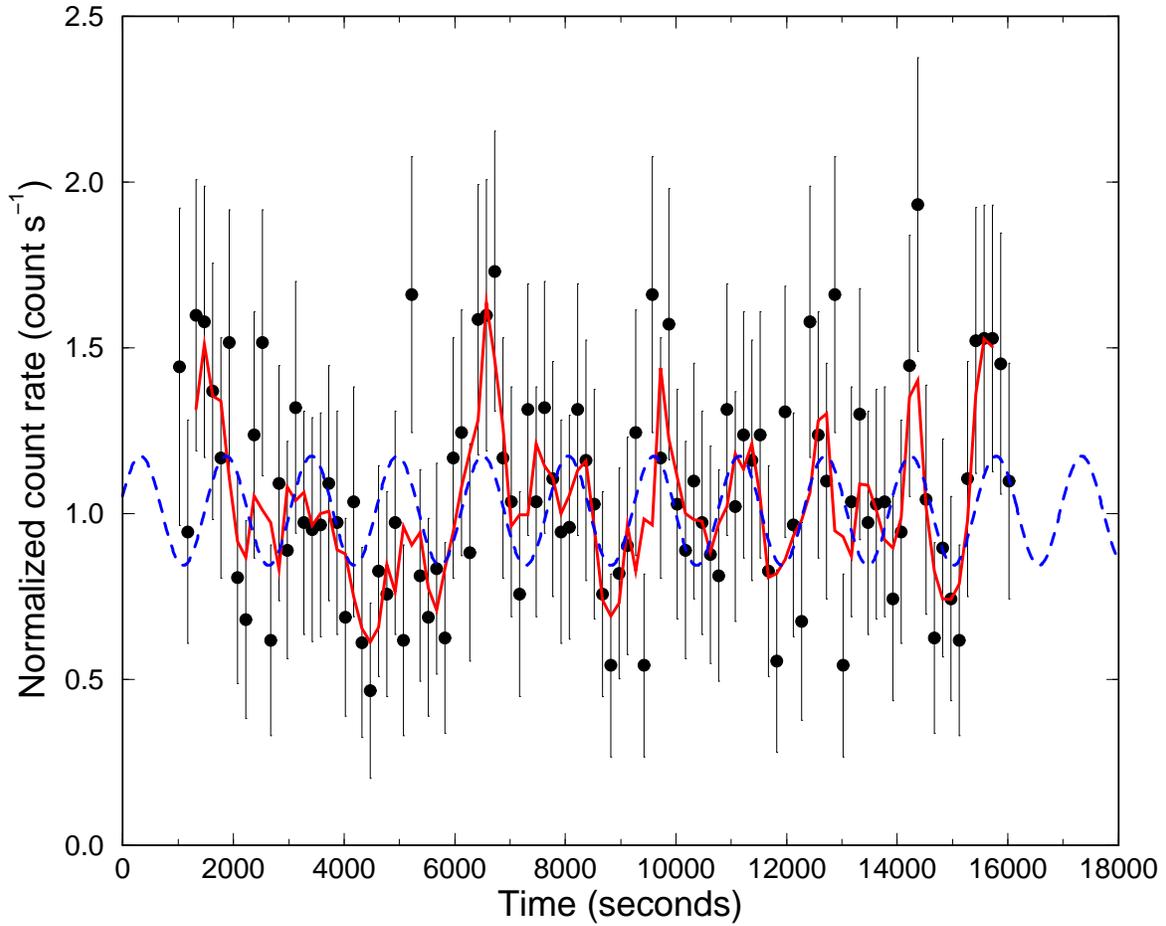}}
\caption{The normalized 1WGA~J1346.5$-$6255 lightcurve obtained with {\it 
Chandra} after background subtraction and binning with 150~s (data points with 1$\sigma$ error bars), 
smoothed with a running window of 450~s (solid 
line) and the sinusoidal fit to the binned data (dashed line). The time 
origin corresponds to 2004 December 26 at 09:22:20~UT (MJD~53365.3905). 
The data show significant variability and a quasi-periodic signal of 
$\sim$1500~s.}
\label{figure:lc_chandra}
\end{center}
\end{figure}

\begin{figure}
\begin{center}
\centerline {\plotone{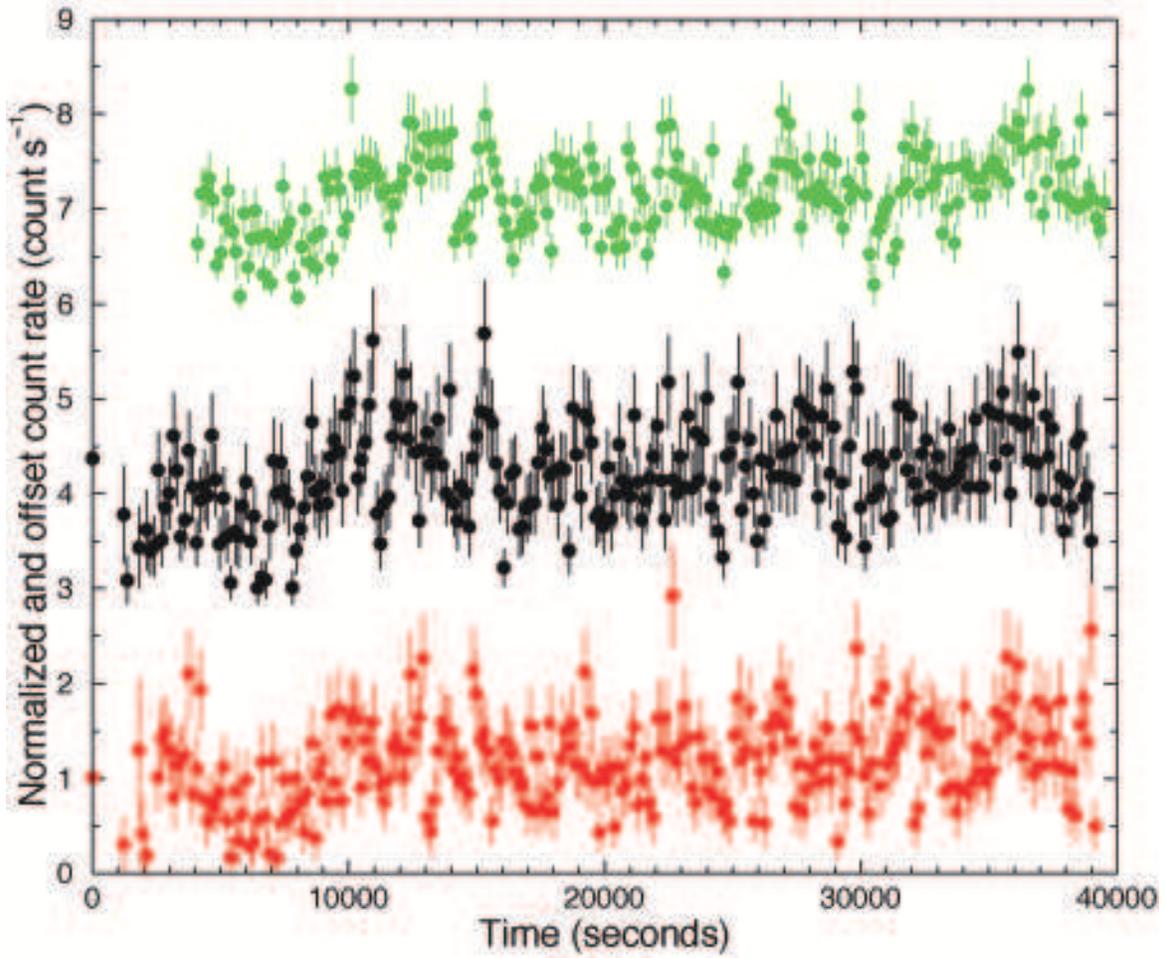}}
%images/wga-lightcurves-v2.eps}}
\caption{The normalized 1WGA~J1346.5$-$6255 lightcurves obtained with {\it 
XMM-Newton} PN (top, green), MOS1 (middle, black), and MOS2 (bottom, red) 
data after background subtraction and binning with 150 seconds. 
Error bars are at the 1$\sigma$ level.
The time origin corresponds to 2001 August 28 
at 02:52:50~UT (MJD~52149.1200). The data show significant variability 
with an amplitude at the 45--55\% level with a timescale of a few 100 seconds 
(see \S4.1 for details).}
%but no clear period is detected.}
\label{figure:lc_xmm}
\end{center}
\end{figure}

We subsequently fitted the spectrum of the diffuse emission from the SNR 
G309.2$-$00.6. While a detailed study of the SNR is beyond the scope of 
this paper, we here present the spectral fit in order to compare its 
column density to that of 1WGA~J1346.5$-$6255 (see 
Fig.~\ref{figure:G309_XMMima}), and therefore test for the association 
between the compact source and the SNR. Since G309.2$-$00.6 is believed to 
be a young SNR \citep{rakowski01}, we fitted its spectrum with a 
non-equilibrium ionization (NEI) model with variable abundances ({\it 
vnei} in XSPEC, \citealt{borkowski01}). A {\it vnei} model with solar 
abundances did not yield an acceptable fit ($\chi^2_{\nu}$ = 10.5, 
$\nu=169$ degrees of freedom), a result that is consistent with the 
previous {\it ASCA} study by \cite{rakowski01}. We subsequently varied the 
abundances of O, Ne, Mg, Si, S, Ar, Ca, Fe and Ni; and found that the 
model requires an over-solar abundance of Si and S, and an under-solar 
abundance of O, Ne, Mg, Ca and Fe. The plasma temperature is $kT$ = 
2.0$\pm$0.6~keV and the column density is $N_{\rm H}$ = 
0.65$^{+0.45}_{-0.25}$ $\times10^{22}$~cm$^{-2}$ (errors here are at the 
3$\sigma$ level, see Fig.~\ref{figure:snr_m1m2pn_vnei_NHkTcont}). This 
best fit $N_{\rm H}$ value is at least three times higher than that 
derived for 1WGA~J1346.5$-$6255 (see Tables~\ref{table:chandra} and 
\ref{table:xmm}). Even considering the upper 3$\sigma$ value for 
1WGA~J1346.5$-$6255 with {\it Chandra} or {\it XMM} and the lower 
3$\sigma$ value for G309.2$-$00.6, $N_{\rm H}$ of the diffuse emission 
from the SNR is still higher. Therefore, based on the $N_{\rm H}$ values 
alone, we can conclude that 1WGA~J1346.5$-$6255 is unlikely associated 
with the SNR. We note that this result does not agree with Rakowski et 
al.'s result, likely due to the lack of resolving power with {\it ASCA} 
which did not allow a study of the SNR independently of the 
1WGA~J1346.5$-$6255 point source.

\section{Optical results} \label{optical}

\begin{figure}
\center
\resizebox{1.0\hsize}{!}{\includegraphics[angle=-90]{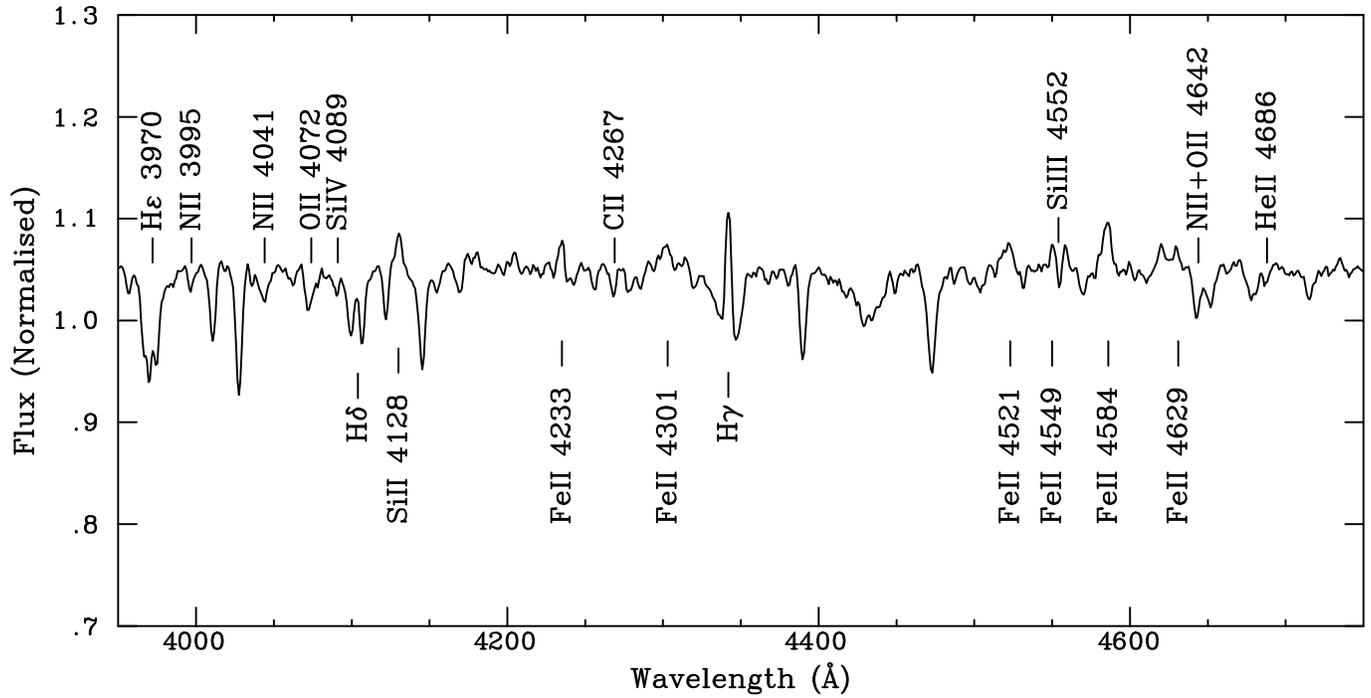}}
%images/optspec.eps}}
\caption{Classification spectrum of HD~119682, taken with EMMI on the NTT 
on 2003 June 5th. The most prominent emission lines and those absorption 
lines useful for classification purposes are labeled. For readability, 
none of the \ion{He}{1} transitions typical of B-type spectra has been 
labeled. The \ion{C}{3}~4650~\AA\ line is not labeled either. 
Surprisingly, it is clearly weaker than the neighboring \ion{O}{2} 
(+\ion{N}{2}) 4642~\AA\ line, suggesting a moderate N enhancement and C 
depletion. A classification of B0.5\,Ve is inferred.}
\label{figure:optspec}
\end{figure}

The normalized classification spectrum of HD~119682 is shown in 
Fig.~\ref{figure:optspec}. It is covered in emission lines, most 
corresponding to \ion{Fe}{2} transitions, and looks typical of Be stars. 
Spectral classification is complicated by the presence of many weak 
emission lines which render identification of weak absorption features 
very difficult. The spectral type can be estimated from the presence of 
both \ion{Si}{3} and \ion{Si}{4} lines and a very weak 
\ion{He}{2}~4686~\AA\ line to be approximately B0.5. The determination of 
the luminosity class is complicated by the emission components and 
broadened lines, but the weakness of \ion{Si}{4}~4089~\AA\ indicates that 
the star is not very luminous and suggests a luminosity class V. This is 
reinforced by the value of $\log g=4.00\pm0.10$~dex found by 
\cite{levenhagen04}, which is consistent with the $\log g=3.9$ values for 
main sequence stars but clearly incompatible with the $\log g=3.5$ values 
for giants listed in \cite{martins05}. Therefore, we will adopt a spectral 
classification of B0.5\,Ve.

We note that the \ion{C}{3}~4650~\AA\ line is abnormally weak for a star 
of this spectral type, as it is always stronger than \ion{O}{2}~4642~\AA\ 
for any star earlier than B2. The \ion{N}{2}~3995, 4041~\AA\ lines are 
moderately stronger than expected, suggesting that there is a certain 
degree of CNO anomaly. This could explain the abnormal 
\ion{C}{3}~4650~\AA/\ion{O}{2}~4642~\AA\ ratio, as C could be moderately 
depleted and the $\lambda$~4642~\AA\ line could have a contribution from 
\ion{N}{2}~4640~\AA.

The Equivalent Width of H$\alpha$ emission line is $-24\pm2$~\AA. The 
shape is typical of a Be star with strong emission, and shows no 
discernible structure or asymmetry.

\section{Discussion} \label{discussion}

\subsection{Astrometry of HD~119682 and 1WGA~J1346.5$-$6255} 
\label{astrometry}

The most precise position and proper motion of HD~11968 are given in 
\cite{zacharias04}. Considering that our {\it Chandra} observations were 
conducted on year 2004.987, we obtain the following J2000 coordinates for 
that epoch: $\alpha=13^{\rm h}~46^{\rm m}~32\fs579\pm0\fs005$, 
$\delta=-62\degr~55\arcmin~24\farcs17\pm0\farcs03$. This is perfectly 
consistent with the {\it Chandra} position of 1WGA~J1346.5$-$6255: 
$\alpha=13^{\rm h}~46^{\rm m}~32\fs6$, 
$\delta=-62\degr~55\arcmin~24\arcsec$ (with 1\farcs3 error radius at 90\% 
confidence level). However, given the high density of stars in the region 
where the X-ray source is located, we have estimated the chance 
coincidence probability of a 2MASS object (assuming negligible error in 
position, typically $\la0\farcs1$) falling within the {\it Chandra} error 
circle in position. We have done this by counting the 2MASS objects 
located within circles of different radius (1, 2, 3, 5, and 10\arcmin) 
around the {\it Chandra} position, and found that the probability is 
always smaller than 4\% for objects brighter than those present in the 
2MASS catalog for this field: $J\simeq18$, $H\simeq17$, and 
$K_s\simeq16$~mag. In fact, the absolute difference in position is just 
0\farcs22, much smaller than the 1\farcs3 {\it Chandra} position error, 
providing a chance coincidence probability of $\simeq0.1$\%, which 
practically confirms without any doubt the association between HD~11968 
and 1WGA~J1346.5$-$6255.

\subsection{Distances to HD~119682/1WGA~J1346.5$-$6255 and SNR G309.2$-$00.6}
\label{distance}

Our spectral type and luminosity class estimates of HD~119682 allow us to 
compute the distance to the star, and hence to 1WGA~J1346.5$-$6255. For a 
B0.5\,V star the absolute magnitude is $M_V=-3.6\pm0.4$~mag (slightly 
extrapolating from \citealt{martins05}) while the intrinsic color is 
$(B-V)_0=-0.28\pm0.02$ \citep{johnson66}. The average of the photometry in 
the literature, reported in \S\ref{introduction}, is $V=7.97\pm0.06$ and 
$B-V=0.16\pm0.05$~mag. The color excess is then $E_{B-V}=0.44\pm0.07$~mag 
and the visual extinction is $A_V=1.41\pm0.24$~mag (with 
$A_V=(3.30+0.28(B-V)_{0}+0.04E_{B-V})\,E(B-V)$ from \citealt{schmidt82}). 
We finally derive a distance modulus of $V-M_V-A_V=10.2\pm0.5$ and a 
distance to the source of $d=1080\pm230$~pc.

However, HD~119682 is in fact a Be star, so the contribution from the 
stellar envelope to $V$ and $E_{B-V}$ should be considered. According to 
\cite{dachs88} and using our measure of the H$\alpha$ Equivalent Width 
($EW$) of $-24\pm2$~\AA, we should subtract a contribution from the 
envelope of $+0.05\pm0.02$~mag to $E_{B-V}$ and $-0.08\pm0.02$~mag to $V$, 
leading to star values of $V=8.05\pm0.06$ and $E_{B-V}=0.39\pm0.07$~mag. 
This provides a visual absorption of $A_V=1.26\pm0.23$~mag, a distance 
modulus of $V-M_V-A_V=10.4\pm0.5$, and a distance of $d=1200\pm260$~pc. We 
note that our measure of the $EW$ of H$\alpha$ is not simultaneous to any 
of the used photometric data. However, since we use the average 
photometry, we expect these corrections to be approximately valid even if 
using a single value of $EW$. This distance estimate to HD~119682, of 
$1.2\pm0.3$~kpc, is in agreement, within errors, with the distance 
estimates to NGC~5281 reported in the literature, the average of which is 
$1.3\pm0.3$~kpc. Therefore, this will be the distance to HD~119682 adopted 
hereafter. We note that this is much lower than the distance estimates to 
SNR~G309.2$-$00.6 appearing in the literature: $5.4\pm1.6$~kpc 
\citep{gaensler98} and $4\pm2$~kpc \citep{rakowski01}.

Finally, by using $A_V=1.26\pm0.23$~mag and the relationship of 
\cite{predehl95} we obtain a hydrogen column density of $N_{\rm 
H}=(2.3\pm0.4)\times10^{21}$~atoms~cm$^{-2}$. This is perfectly 
consistent, within errors, with the values obtained from the X-ray 
spectral fitting of the {\it Chandra} and {\it XMM-Newton} data (see 
Figs.~\ref{figure:Chandra_powerlaw_cont} and 
\ref{figure:XMM_powerlaw_cont} and Tables~\ref{table:chandra} and 
\ref{table:xmm}). Therefore, there is no need for intrinsic absorption 
around the X-ray emitting zone.

We can now estimate the distances to the X-ray sources based on the 
measurements of the hydrogen column density, their conversion to 
extinction according to \cite{predehl95} ( $<N_H/E_{B-V}>$ = 5.55 $\times$ 
10$^{21}$~cm$^{-2}$~mag$^{-1}$), and the value of the extinction per unit 
distance in the direction of the sources, which is of 
$E_{B-V}/d\sim0.3$~mag~kpc$^{-1}$ \citep{lucke78}. For this purpose we use 
the most accurate values of $N_{\rm H}$, obtained with the {\it 
XMM-Newton} data. For 1WGA~J1346.5$-$6255 we consider the power-law model 
fit with $N_{\rm H}\sim(2.2\pm0.4)\times10^{21}$~atoms~cm$^{-2}$ 
(3$\sigma$ uncertainties) and obtain a distance of $d=1.3\pm0.3$~kpc, 
fully consistent with the distance to HD~119682 and NGC~5281. (We note 
that the column density derived for the best fit power-law+MEKAL or 
power-law+Gaussian is close enough to that derived using the power-law and 
thus yields a similar distance estimate within error). For the diffuse 
emission of SNR G309.2$-$00.6 we use the {\it vnei} model fit with $N_{\rm 
H}=6.5^{+4.5}_{-2.5}\times10^{21}$~atoms~cm$^{-2}$ (3$\sigma$ 
uncertainties) and obtain a distance of $d\la3.9^{+2.4}_{-1.5}$~kpc, 
consistent with the $4\pm2$~kpc value obtained by \cite{rakowski01} from 
their {\it ASCA} fit to the SNR. This distance is clearly incompatible 
with the distance to HD~119682/1WGA~J1346.5$-$6255, but consistent within 
errors with the distance of $5.4\pm1.6$~kpc to the RCW~80 \ion{H}{2} 
region proposed by \cite{gaensler98} to be associated with SNR 
G309.2$-$00.6.

\subsection{Circumstellar envelope} \label{envelope}

The intrinsic $(J-K_s)_0$ infrared color of a normal B0,1\,V star is 
$-0.22$ \citep{ruelas91}. The 2MASS photometry provides a value of 
$J-K_s=0.40\pm0.04$, leading to a total infrared color excess of 
$E_{J-K_s}=0.62\pm0.04$. On the other hand, from the visual absorption of 
interstellar origin found above, $A_{V,~\rm IS}=1.26\pm0.23$~mag, and 
using the relationships by \cite{rieke85} we found interstellar infrared 
absorptions of $A_{J,~\rm IS}=0.36\pm0.06$ and $A_{K_s,~\rm 
IS}=0.14\pm0.03$, and finally $E_{J-K_s,~\rm IS}=0.22\pm0.07$. Therefore, 
in addition to the interstellar color excess, there is an extra 
contribution to $E_{J-K_s}$ of $0.40\pm0.08$. This is typical of Be stars 
with strong emission lines, usually adscribed to their circumstellar 
envelope.

\subsection{Age of HD~119682} \label{age}

\cite{levenhagen04} report an age of $4\pm1$~Myr and a mass of 
$M=18\pm1~M_\odot$ for HD~119682 based on the evolutionary tracks by 
\cite{schaller92}. In fact, according to the new evolutionary tracks by 
\cite{meynet03} (see their Fig.~5), the luminosity and effective 
temperature reported by \cite{levenhagen04} are in agreement with a 
18~$M_\odot$ star at the beginning of the main sequence or with a slightly 
evolved 20~$M_\odot$ star. However, since new calibrations such as the one 
of \citealt{martins05} favor lower masses, we will consider as a strict 
upper limits for the age of HD~119682 the main sequence lifetime of a 
15~$M_\odot$ star. On the other hand, since HD~119682 has a high 
rotational velocity of $v \sin i = 220\pm20$~km~s$^{-1}$, we will use the 
evolutionary tracks by \cite{meynet03} for an initial rotational velocity 
of 300~km~s$^{-1}$, leading to a total main sequence lifetime of 14.5~Myr 
(see their Table~1). Therefore, even if assuming the lowest possible mass 
and a high rotational velocity, the lifetime in the main sequence is less 
than 15~Myr, i.e., 3 times lower than the estimated age of NGC~5281 by 
\cite{sanner01}. This clearly confirms the blue straggler nature of 
HD~119682, which is not clear in the $V$ vs. $(B-V)$ CMD of 
\cite{sanner01}, slightly affected by the reddening of the envelope, but 
clearly evident when looking at the $V$ vs. $(U-B)$ CMD of 
\cite{moffat73}.

\subsection{On the nature of HD~119682/1WGA~J1346.5$-$6255} \label{nature}

The unabsorbed X-ray luminosity of 1WGA~J1346.5$-$6255 from our {\it 
Chandra} observations is $L_{(0.5-7.5~{\rm 
keV})}=2.5^{+0.6}_{-0.7}\,(d/1.3~{\rm kpc})^2\times10^{32}$~ergs~s$^{-1}$ 
(using the range of fluxes tabulated in Table~\ref{table:chandra}). From 
the {\it XMM-Newton} observations and using the range of fluxes obtained 
for the various models (Table~\ref{table:xmm}), we obtain a luminosity of 
$L_{(0.5-8.5~{\rm keV})}=4.5^{+0.1}_{-0.7}\,(d/1.3~{\rm 
kpc})^2\times10^{32}$~ergs~s$^{-1}$, which corresponds to 
$L_{(0.5-7.5~{\rm keV})}=4.1^{+0.2}_{-0.6}\,(d/1.3~{\rm 
kpc})^2\times10^{32}$~ergs~s$^{-1}$, about a factor of 1.5--2 higher than 
the {\it Chandra} value, regardless of the model used. In any case, this 
is clearly lower than that of Be/X-ray binaries containing neutron stars, 
even the faint persistent ones (which have $L_{\rm 
X}\sim10^{34}$~ergs~s$^{-1}$, see \citealt{reig99}). Moreover, no X-ray 
pulsations have ever been detected, although \cite{rakowski01} could only 
constrain the pulsed fraction to be less than $\sim$85\% in the frequency 
range 0.002--32~Hz (0.03--500~s) at a 99.99\% confidence level. Therefore, 
the currently available information does not favor the hypothesis of a 
binary system with an accreting neutron star as the compact object.

Another possibility would be to have a relatively quiescent accreting 
black hole in a binary system. Indeed, the photon index of 
1WGA~J1346.5$-$6255 is similar to those found in accreting black holes 
while in the low/hard state (see \citealt{fender04} and references 
therein). \cite{gallo03} found a correlation between the 2--11~keV X-ray 
flux and the cm radio flux density for black holes in the low/hard state. 
>From our power-law model fit to the {\it XMM-Newton} data, we obtain an 
unabsorbed flux of $F_{\rm 2-11~keV}= 
1.7\times10^{-12}$~ergs~cm$^{-2}$~s$^{-1}$ and compute an expected radio 
flux density of $0.29\pm0.20$~mJy at a distance of $1.3$~kpc. Given the 
uncertainties, such a source could be above or below the 3$\sigma$ level 
in the ATCA image of \cite{gaensler98}. Deeper radio observations, with a 
noise smaller than 0.02~mJy, and with long baselines to avoid 
contamination from the SNR, would be needed to properly search for a radio 
counterpart. Only simultaneous X-ray observations could be useful to 
exclude the black hole scenario, although this would also be uncertain due 
to the scatter of the radio/X-ray correlation. However, we note that there 
is no clear evidence that a Be+black hole binary system has been found up 
to now. In addition, the X-ray spectra are better fitted including one or 
several MEKAL models, which is not the case for black holes in the 
low/hard state. Therefore, the scenario of an accreting black hole 
orbiting the Be star is not favored by the available data.

OB stars are known to be X-ray sources, presumably due to shocks arising 
in their stellar winds (see \citealt{guedel04} for a review). Based on 
{\it Einstein} data, \cite{pallavicini81} found that the X-ray 
luminosities of $\sim$30 OB stars followed approximately the empirical law 
$L_{\rm X}\simeq(1.4\pm0.3)\times10^{-7}L_{\rm bol}$. A more complete 
study based on {\it ROSAT} data for more than 200 OB stars allowed 
\cite{berghoefer97} to propose the existence of a correlation with two 
different power laws above and below $L_{\rm bol}=10^{38}$~ergs~s$^{-1}$. 
The $\log L/L_\odot=4.64\pm0.10$ value given by \cite{levenhagen04} 
provides a star bolometric luminosity of $L_{\rm 
bol}=(1.7\pm0.4)\times10^{38}$~ergs~s$^{-1}$, leading, according to 
\cite{berghoefer97}, to an X-ray luminosity of $L_{\rm 
X}=(2^{+3}_{-1})\times10^{31}$~ergs~s$^{-1}$. In contrast, the unabsorbed 
power-law X-ray luminosity from our {\it Chandra} observations, 
$L_{(0.5-7.5~{\rm keV})}=2.5^{+0.6}_{-0.7}\,(d/1.3~{\rm 
kpc})^2\times10^{32}$~ergs~s$^{-1}$ is around one order of magnitude 
higher, while the {\it XMM-Newton} luminosity is even a factor of about 2 
higher. In fact, the X-ray emission of HD~119682 is not only stronger than 
expected, but obviously harder than that of isolated stars: our {\it 
XMM-Newton} two-component MEKAL fit to the data requires a high 
temperature of $kT=13.0^{+2.6}_{-2.4}$~keV 
(14.3$^{+2.9}_{-2.6}\times10^7$~K), more than an order of magnitude higher 
than usual in main sequence B stars, and a factor of at least four higher 
than the most extreme cases \citep{cohen97}.

As a matter of fact, the luminosity and spectral shape of 
HD~119682/1WGA~J1346.5$-$6255 are similar to those displayed by the 
notoriously peculiar Be star $\gamma$~Cas: $L_{(2-10~{\rm 
keV})}=6\times10^{32}$~ergs~s$^{-1}$ and two-component MEKAL spectrum with 
temperatures of $0.05\pm0.01$~keV and $12.3\pm0.6$~keV \citep{owens99}. 
Recently, \cite{smith06} have proposed that the star HD~110432, which 
shows $L_{(2-10~{\rm keV})}\simeq5\times10^{32}$~ergs~s$^{-1}$ and a MEKAL 
spectrum with $kT=10.6\pm1.9$~kev \citep{torrejon01}, is an almost perfect 
twin of $\gamma$~Cas and have mentioned the possibility of the existence 
of a class of objects displaying similar characteristics, which they dub 
``$\gamma$~Cas analogs''. In addition, \cite{motch05} have very recently 
presented a summary of X-ray and optical properties of their proposed 
currently existing five ``$\gamma$~Cas-like objects'' (plus $\gamma$~Cas 
itself). The properties of HD~119682/1WGA~J1346.5$-$6255, to be compared 
with their Table~1, are: spectral type B0.5\,Ve, H$\alpha$ EW of 
$-$24~\AA, $L_{\rm X~(0.2-12~keV)}=3.6$--$6.6\times10^{32}$~ergs~s$^{-1}$ 
(from the {\it Chandra} and {\it XMM-Newton} power-law model fits, 
respectively), $kT_{\rm soft}=1.7$~keV and $kT_{\rm hard}=13$~keV, and 
$\Gamma\sim$1.7. These properties are practically identical to those of 
SS~397, with a slightly different equivalent width of H$\alpha$. 
Furthermore, one of the objects in the list of \cite{motch05} is a blue 
straggler in the 50~Myr old open cluster NGC~6649, closely matching the 
observed properties of HD~119682, also a blue straggler and located in the 
$\sim$45~Myr old open cluster NGC~5281.

In the previous spectroscopic study, we showed that the X-ray flux of the 
1WGA~J1346.5$-$6255 source is a factor of $\sim$2 brighter with the {\it 
XMM} observation than with the {\it Chandra} one, obtained 3 years later. 
This could be attributed to intrinsic variability of the source. The 
short-term variability detected on timescales of a few hundred seconds is 
similar to that seen in other $\gamma$~Cas-like objects \citep{motch05}. 
Moreover, the periodicity of $\sim$1500~s we have detected in 
HD~119682/1WGA~J1346.5$-$6255, with a pulse fraction of 13--16\%, is 
reminiscent to, although not so clear than, the oscillation found in the 
$\gamma$~Cas analog HD~161103, with a period of $3250\pm350$~s and a pulse 
fraction of 24\% \citep{lopes06}. However, in the case of 
HD~119682/1WGA~J1346.5$-$6255, this possible period has been detected in 
two different observations performed with two different satellites, 
indicating that this signal is probably stable along time (although other 
superimposed signals are detected in the {\it XMM-Newton} data set).

In summary, HD~119682/1WGA~J1346.5$-$6255 shares the following properties 
with the other $\gamma$~Cas analogs: spectral fits including thermal 
components are better than simple power-law fits (see 
Tables~\ref{table:chandra} and \ref{table:xmm}); in the two-temperature 
MEKAL model fits, the cooler component is much fainter than the hard one 
(see Tables~\ref{table:chandra} and \ref{table:xmm}); the total 
photoelectric absorption in X-rays is not very different from that due to 
the interstellar medium; the X-ray luminosity is in the range $L_{\rm 
X~(0.2-12~keV)}=10^{32}$--$10^{33}$~ergs~s$^{-1}$; there are no reported 
X-ray outbursts; the flux varies by a factor of a few from one observation 
to the other; variability on short timescales is detected, showing 
different quasi-periodic signals; the presence of an iron line near 
$\sim$6.7~keV (see \S4.1 and Fig.~\ref{figure:XMM_powerlaw_fit}), 
difficult to interpret in the black hole scenario; a moderate N 
enhancement; and a circumstellar envelope that contributes to the NIR 
reddening.

Therefore, we conclude that the overall properties of HD~119682 indicate 
that it is most likely a new $\gamma$~Cas analog.

The nature of $\gamma$~Cas analogs is still a matter of debate (see 
\citealt{motch05} and \citealt{lopes06} for recent discussions). Possible 
scenarios to explain the properties of the detected X-ray emission are the 
magnetic interaction between the star and its circumstellar decretion 
disk, or accretion on to a compact object (neutron star, white dwarf, or 
even a black hole). We note that the possible $\sim$1500~s period we have 
detected in HD~119682/1WGA~J1346.5$-$6255 could be the rotational period 
in the neutron star or white dwarf scenario (although a long observation 
with {\it XMM-Newton} is needed to confirm it). In principle, optical 
spectroscopic observations of HD~119682 could unveil the presence of a 
compact object by means of a radial velocity curve, as has been done in 
the the case of $\gamma$~Cas itself \citep{harmanec00,miroshnichenko02}. 
However, Be stars in X-ray binaries typically have orbital periods in the 
range $\sim$10--200~d, so a long-lasting observing program would be 
needed. In addition, the number and intensity of emission lines along the 
whole spectrum would render difficult to obtain accurate radial velocity 
measurements. Certainly, to obtain a radial velocity curve is not a 
straightforward task in the case of HD~119682.

Whatever the nature of HD~119682/1WGA~J1346.5$-$6255, we stress that it is 
not related to the background SNR G309.2$-$00.6.

\section{Summary} \label{summary}

After a study of the sources 1WGA~J1346.5$-$6255 and SNR G309.2$-$00.6 
with {\it Chandra} and {\it XMM-Newton} and of HD~119682 with the NTT, we 
have obtained the following results:

\begin{itemize}

\item{The 1WGA~J1346.5$-$6255 source is consistent with a point source. No 
jets have been observed with {\it Chandra}. We derive an upper limit on 
the unabsorbed flux from any unseen jets of 2.6$\times$10$^{-13}$ 
ergs~cm$^{-2}$~s$^{-1}$ (0.5--7.5~keV), which corresponds to a luminosity 
of 5.3$\times$10$^{31}$~ergs~s$^{-1}$ at a distance of 1.3~kpc.}

\item{The X-ray spectrum of 1WGA~J1346.5$-$6255 is best fitted with 
two-component models including a thermal component. A two-component 
power-law+MEKAL model provides the best fits to the {\it Chandra} and {\it 
XMM} data. The hard component dominates the X-ray emission.}

\item{The {\it Chandra} and {\it XMM}-derived $N_{\rm H}$ of 
(1--3)$\times$10$^{21}$~cm$^{-2}$ (conservative range encompassing all 
models used) is lower than that towards the SNR: 
(4--11)$\times$10$^{21}$~cm$^{-2}$ (3$\sigma$ range).}

\item{The {\it Chandra} flux of $\sim$ 1.0--1.5 
$\times$10$^{-12}$~ergs~cm$^{-2}$~s$^{-1}$ (0.5--7.5~keV, unabsorbed) is a 
factor of 1.5--2 lower than that observed with {\it XMM-Newton}. 
Variability on short timescales (a few 100 seconds) has been also detected 
at a level of 45--55\%, with a possible period of $\sim$1500~s in the {\it 
Chandra} and {\it XMM} data sets, which could be the rotational period of 
a neutron star or a whide dwarf.}

\item{1WGA~J1346.5$-$6255 is coincident with the optical star HD~119682 
whose spectrum is that of a B0.5\,Ve star. It is a blue straggler in the 
$\sim$45~Myr old open cluster NGC~5281. The distance is 1.3$\pm$0.3~kpc, 
significantly lower than the distance to the SNR G309.2$-$00.6.}

\item{Based on the above optical and X-ray spectral and timing properties 
of 1WGA~J1346.5$-$6255, we conclude that it is unlikely to be a 
microquasar associated with the SNR G309.2$-$00.6, and that it is most 
likely a new member of the growing class of $\gamma$-Cas objects.}

\item{Further X-ray observations are needed to study the variability of 
the source, and further explore the bump near 6.7~keV. Such a study will 
confirm the nature of 1WGA~J1346.5$-$6255 as a $\gamma$-Cas like object.}

\item{The search for a compact object associated with the G309.2$-$00.6 
requires a deep {\it Chandra} observation of the SNR.}

\end{itemize}

\acknowledgements

Partly based on observations collected at the European Southern
Observatory, La Silla, Chile (observing proposal ESO N$^{\rm o}$ 071.D-0151).
S.S.-H. and H. M. acknowledge 
support by the Natural Sciences and Engineering Research
Council (NSERC) of Canada. Partial support was also provided by
NASA grant NNG05GL15G.
This research is also supported by the Spanish Ministerio de Educaci\'on y
Ciencia (MEC) through grants AYA2004-07171-C02-01 and AYA2005-00095, 
partially funded by the European Regional Development Fund (ERDF/FEDER).
M.R. has been supported by the French Space Agency (CNES)
and by a Marie Curie Fellowship of the European Community programme
Improving Human Potential under contract number HPMF-CT-2002-02053, and is
being supported by a {\em Juan de la Cierva} fellowship from MEC.
I.N. is a researcher of the programme {\em Ram\'on y Cajal}, funded by the
Spanish Ministerio de Educaci\'on y Ciencia and the University of Alicante,
with partial support from the Generalitat Valenciana and the European Regional
Development Fund (ERDF/FEDER).
Y.M.B. acknowledges the support of a NASA Long Term Space Astrophysics grant.
This research has made use of NASA's Astrophysics Data System Abstract 
Service, the HEASARC database operated by NASA's Goddard Space Flight 
Center, and of the SIMBAD database operated by the CDS, Strasbourg, 
France.
This publication made use of data products from the Two Micron All Sky 
Survey, which is a joint project of the University of Massachusetts and 
the Infrared Processing and Analysis Center/California Institute of 
Technology, funded by the National Aeronautics and Space Administration 
and the National Science Foundation.

%figure 1

%figure 2

%figure 3

%figure 4

%figure 5

%figure 6

%figure 7

%figure 8

%figure 9

%figure 10

%figure 11

%figure 12

%figure 13

%figure 14

%figure 15

%figure 16

\renewcommand{\baselinestretch}{1}

\clearpage

\begin{deluxetable}{ccc}
\tablecaption{Photometry of HD~119682.
\label{table:optical}}
\tablewidth{0pt}
\tablehead{\colhead{Ref.} & \colhead{$V$ (mag)} & \colhead{$B-V$ (mag)}}
\startdata
\cite{moffat73}    & $7.98\pm0.01$     & $0.13\pm0.01$     \\
\cite{humphreys75} & $8.00$            & $0.20$            \\
\cite{drilling91}  & $8.06\pm0.01$     & $0.22\pm0.01$     \\
\cite{hog00}       & $7.90\pm0.01$     & $0.08\pm0.02$     \\
\cite{sanner01}    & $7.91\pm0.04$     & $0.16\pm0.06$     \\
\hline
Weighted average   & $7.97\pm0.06$     & $0.16\pm0.05$     \\
\enddata
\tablecomments{The Tycho-2 magnitudes \citep{hog00} have been converted from
$VT$ and $BT$ to $V$ and $B-V$ magnitudes following their prescription. We
have assumed an error of 0.05~mag for the \cite{humphreys75} photometry to
compute the weighted average. Errors are at the 1$\sigma$ level.}
\end{deluxetable}

\clearpage

\begin{deluxetable}{ccc}
\tabletypesize{\small}
\tablecaption{Spectral fit parameters of 1WGA~J1346.5$-$6255 obtained with {\it
Chandra}.
\label{table:chandra}}
\tablewidth{0pt}
\tablehead{\colhead{Model} & \colhead{Parameters} & \colhead{Fitted values}}
\startdata
wabs*power &
$N_{\rm H} (\times 10^{22}~$cm$^{-2})$ & 0.18 (0.11--0.26)\\
& $\Gamma$ & 1.41 (1.19--1.65)\\
& norm $(\times 10^{-4})$ & 1.58 (1.28--1.96)\\
& $\chi_\nu^2$ ($\nu$) & 0.463 (61)\\
& flux $(\times 10^{-12}$~ergs~cm$^{-2}$~s$^{-1})$ & 1.13 (0.89--1.42)\\
\hline
wabs*mekal &
$N_{\rm H} (\times 10^{22}~$cm$^{-2})$ & 0.15 (0.10--0.21)\\
& $kT$ (keV) & 26.9 (9.7--79.9) \\
& norm $(\times 10^{-4})$ & 7.02 (5.92--9.16)\\
& $\chi_\nu^2$ ($\nu$) & 0.480 (61)\\
& flux $(\times 10^{-12}$~ergs~cm$^{-2}$~s$^{-1})$ & 1.13 (0.94--1.23)\\
\hline
wabs*(mekal+mekal) &
$N_{\rm H} (\times 10^{22}~$cm$^{-2})$ & 0.19 (0.12--0.28)\\
& $kT_1$ (keV) & 0.95 (0.71--1.29)\\
& norm$_1$ $(\times 10^{-5})$ & 3.10 (1.26--5.66)\\
& $kT_2$ (keV) & 80 ($\geq$14) \\
& norm$_2$ $(\times 10^{-4})$ & 8.2 (5.9--9.4)\\
& $\chi_\nu^2$ ($\nu$) & 0.342 (59)\\
& flux $(\times 10^{-12}$~ergs~cm$^{-2}$~s$^{-1})$ & 1.16 (1.10--1.23)\\
\hline
wabs*(power+mekal) & 
$N_{\rm H} (\times 10^{22}~$cm$^{-2})$ & 0.19 (0.11--0.32)\\
& $\Gamma$ & 1.24 (0.98--1.52)\\
& norm$_\Gamma$ $(\times 10^{-4})$ & 1.30 (0.98--1.53)\\
& $kT$ (keV) & 0.97 (0.69--1.34) \\
& norm$_{kT}$ $(\times 10^{-5})$ & 3.38 (1.09--5.85)\\
& $\chi_\nu^2$ ($\nu$) & 0.341 (59)\\
& flux $(\times 10^{-12}$~ergs~cm$^{-2}$~s$^{-1})$ & 1.18 (0.89--1.53)\\
\enddata
\tablecomments{All fits have been performed in the 0.5--7.5~keV energy 
range. The quoted fluxes are unabsorbed in the same energy range. The 
MEKAL models assume solar abundances. All confidence ranges are 90\%.}
\end{deluxetable}

\clearpage

\begin{deluxetable}{ccc}
\tabletypesize{\small}
\tablecaption{Spectral fit parameters of 1WGA~J1346.5$-$6255 obtained with 
{\it XMM-Newton}.
\label{table:xmm}}
\tablewidth{0pt}
\tablehead{\colhead{Model} & \colhead{Parameters} & \colhead{Fitted values}}
\startdata
wabs*power &
$N_{\rm H} (\times 10^{22}~$cm$^{-2})$ & 0.22 (0.20--0.24)\\
& $\Gamma$ & 1.69 (1.64--1.73)\\
& norm $(\times 10^{-4})$ & 3.77 (3.54--3.98)\\
& $\chi_\nu^2$ ($\nu$) & 1.04 (418)\\
& flux $(\times 10^{-12}$~ergs~cm$^{-2}$~s$^{-1})$ & 2.22 (2.20--2.24)\\
\hline
wabs*mekal &
$N_{\rm H} (\times 10^{22}~$cm$^{-2})$ & 0.15 (0.14--0.16)\\
& $kT$ (keV) & 9.2 (8.5--10.8)\\
& norm $(\times 10^{-3})$ & 1.15 (1.11-1.19)\\
& $\chi_\nu^2$ ($\nu$) & 1.12 (418)\\
& flux $(\times 10^{-12}$~ergs~cm$^{-2}$~s$^{-1})$ & 2.15 (2.0--2.18)\\
\hline
wabs*(mekal+mekal) &
$N_{\rm H} (\times 10^{22}~$cm$^{-2})$ & 0.16 (0.15--0.17)\\
& $kT_1$ (keV) & 1.7 (1.4--2.0)\\
& norm$_1$ $(\times 10^{-4})$ & 1.07 (0.42--3.49)\\
& $kT_2$ (keV) & 13.0 (10.6--15.6) \\
& norm$_2$ $(\times 10^{-3})$ & 1.08 (0.9--1.12)\\
& $\chi_\nu^2$ ($\nu$) & 1.025 (416)\\
& flux $(\times 10^{-12}$~ergs~cm$^{-2}$~s$^{-1})$ & 2.16 (2.15--2.17)\\
\hline
wabs*(power+mekal) & 
$N_{\rm H} (\times 10^{22}~$cm$^{-2})$ & 0.21 (0.19--0.24)\\
& $\Gamma$ & 1.60 (1.53--1.67)\\
& norm$_\Gamma$ $(\times 10^{-4})$ & 3.3 (1.6--3.0)\\
& $kT$ (keV) & 1.4 (1.1--2.2) \\
& norm$_{kT}$ $(\times 10^{-5})$ & 4.6 (2.5--16.8)\\
& $\chi_\nu^2$ ($\nu$) & 1.005 (416)\\
& flux $(\times 10^{-12}$~ergs~cm$^{-2}$~s$^{-1})$ & 2.20 (2.18--2.22)\\
\hline
wabs*(power+Gaussian) &
$N_{\rm H} (\times 10^{22}~$cm$^{-2})$ & 0.23 (0.21--0.24)\\
& $\Gamma$ & 1.71 (1.66--1.76)\\
& norm $(\times 10^{-4})$ & 3.85 (3.63--3.97)\\
& $E_{\rm line}$ (keV) & 6.77 (6.65--6.91) \\
& $\sigma_{\rm line}$ (keV) & 0.19 (6$\times$10$^{-3}$--0.32) \\
& norm$_{\rm line}$ $(\times 10^{-6}$) & 4.33 (2.39--6.87)\\
& $\chi_\nu^2$ ($\nu$) & 1.012 (415) \\
& flux $(\times 10^{-12}$~ergs~cm$^{-2}$~s$^{-1})$ & 2.20 (1.88--2.26)\\
\enddata
\tablecomments{All fits have been performed in the 0.5--8.5~keV energy 
range using PN, MOS1 and MOS2 data. The quoted fluxes are unabsorbed in 
the same energy range. The MEKAL models assume solar abundances. All 
confidence ranges are 90\%.}
\end{deluxetable}

\end{document}